%% file: main.tex
\documentclass[11pt]{article}
\usepackage[utf8]{inputenc}
\usepackage[T1]{fontenc}
\usepackage{lmodern}
\usepackage{amsmath,amssymb,amsthm,mathtools}
\usepackage[margin=1in]{geometry}
\usepackage[round,authoryear]{natbib}
\usepackage{graphicx}
\usepackage{booktabs}
\usepackage{array}
\usepackage{enumitem}
\usepackage{microtype}
\usepackage{xcolor}
\usepackage[colorlinks=true,linkcolor=blue!50!black,citecolor=blue!50!black,urlcolor=blue!50!black]{hyperref}
\usepackage{caption}
\usepackage[section]{placeins}
\usepackage{algorithm}
\usepackage{algpseudocode}

\newtheorem{theorem}{Theorem}
\newtheorem{proposition}{Proposition}
\newtheorem{lemma}{Lemma}
\newtheorem{corollary}{Corollary}
\theoremstyle{definition}
\newtheorem{remark}{Remark}

\newcommand{\R}{\mathbb{R}}
\newcommand{\N}{\mathrm{N}}
\newcommand{\Ga}{\mathrm{Gamma}}
\newcommand{\GIG}{\mathrm{GIG}}
\newcommand{\rank}{\operatorname{rank}}
\newcommand{\tr}{\operatorname{tr}}
\newcommand{\diag}{\operatorname{diag}}
\newcommand{\Var}{\operatorname{Var}}
\newcommand{\E}{\mathrm{E}}
\newcommand{\norm}[1]{\lVert #1\rVert}
\newcommand{\Deta}{D_\eta}
\newcommand{\pe}{\preceq}
\newcommand{\ind}{\stackrel{\text{ind}}{\sim}}
\newcommand{\iid}{\stackrel{\text{iid}}{\sim}}

\title{Geometric ergodicity of the block Gibbs sampler for Bayesian linear mixed models with scale mixtures of normal random effects}
\author{Rolando de la Cruz\thanks{Faculty of Engineering and Sciences, Universidad Adolfo Ib\'a\~nez, Pe\~nalol\'en, Chile; Data Observatory Foundation, Santiago, Chile; Millennium Nucleus for Social Data Science (SODAS), Santiago, Chile. Corresponding author: \texttt{rolando.delacruz@uai.cl}}
\and Nathalie Humeniy\thanks{Departamento de Estad\'istica, Pontificia Universidad Cat\'olica de Chile, Santiago, Chile.}}
\date{November 2026}

\begin{document}
\maketitle

\begin{abstract}
We study a Bayesian linear mixed model whose random effects follow a scale mixture of normal distributions with an arbitrary mixing distribution, a class that contains the Student-$t$, slash, contaminated normal, normal-generalized inverse Gaussian and Laplace laws. Conditionally conjugate priors lead to a three-block Gibbs sampler that alternates between the variance components, the latent precision weights, and the regression coefficients together with the random effects. Using a drift function built from fractional powers, we give sufficient conditions for geometric ergodicity of this chain, which involve a moment of the mixing distribution of some order larger than $1/2$ and its behaviour near the origin. An equivariance of the sampler under rescaling of the design then shows that the conditions hold for all the families above, including Student-$t$ random effects with any degrees of freedom, for every random-effects design matrix, full rank or not, with the rank entering only the conditions on the priors of the variance components. Together with posterior moment bounds, geometric ergodicity justifies Monte Carlo standard errors and effective-sample-size stopping rules for the regression coefficients and the log-variance components. Simulations examine these procedures at practical run lengths: joint confidence regions need a correction for the number of batches, and under sequential stopping their observed coverage remains slightly below the nominal level. We illustrate the methods with longitudinal $\beta$-HCG measurements from early pregnancies, in which the latent weights reveal larger between-subject variability among abnormal pregnancies.
\end{abstract}


\input{sec1_intro}
\input{sec2_model}
\input{sec3_results}
\input{sec4_proofs}
\input{sec5_practice}
\input{sec6_sim}
\input{sec7_app}
\input{sec8_disc}

\section*{Code and data}
Python code implementing Algorithm~\ref{alg:gibbs}, the verification tests of Appendix~\ref{app:code}, the simulation studies of Section~\ref{sec:sim} and the analyses of Section~\ref{sec:app}, with the random seeds used and instructions to reproduce every table and figure, is available at \url{https://github.com/rolandodelacruz74/smnlmm-gibbs}. The $\beta$-HCG data of Section~\ref{sec:app} are available from the first author upon request.

\section*{Acknowledgements}
{\sloppy The work of the first author was partially supported by ANID, Chile, through grants ANID/FONDECYT/\allowbreak 1260228, ANID/ANILLO/\allowbreak ACT210096 and ANID-MILENIO-\allowbreak NCN2024\_103, and by the Data Observatory Foundation, ANID Technology Center No.~DO210001. The work of the second author was partially supported by an ANID PhD scholarship.\par}

\appendix
\input{appendix}

\bibliographystyle{plainnat}
\bibliography{paper}
\end{document}

%% file: sec1_intro.tex
\section{Introduction}\label{sec:intro}

Linear mixed models usually assume normally distributed random effects. The assumption is convenient, but departures from normality can affect several aspects of inference. Fixed-effect estimates are often robust to misspecification of the random-effects distribution \citep{Verbeke1997,McCulloch2011SS}, but variance components, predictions of subject-specific effects, likelihood-ratio tests and interval estimates can be seriously affected, particularly when the true distribution has heavier tails than the normal \citep{Hui2021,McCulloch2011Bcs,Drikvandi2017}. A natural and widely used remedy replaces the normal distribution by a scale mixture of normals, also called a normal/independent distribution \citep{Lange1989,Pinheiro2001,Rosa2003,Lachos2010,Schumacher2021}. Each random effect is then normal given a latent precision weight, and the weights downweight subjects whose effects are far from the bulk of the population. The Student-$t$, slash and contaminated normal distributions are the most common choices.

Bayesian inference for these models is carried out with Markov chain Monte Carlo (MCMC). Because the random effects are conditionally normal given the weights, a conditionally conjugate prior leads to a simple Gibbs sampler with three blocks: the variance components, the latent weights, and the vector formed by the regression coefficients and the random effects. This sampler is easy to implement \citep{Rosa2003}. The reliability of the resulting estimates, however, depends on how fast the Markov chain converges. Harris ergodicity ensures that ergodic averages are strongly consistent, but it does not guarantee a central limit theorem (CLT), and without a CLT one cannot attach valid Monte Carlo standard errors to the estimates \citep{fleg:hara:jone:2008,jone:qin:2022}. The standard route to a CLT is to show that the chain is geometrically ergodic. Geometric ergodicity, together with moment conditions, also yields consistent batch-means estimators of the asymptotic covariance matrix and therefore justifies the multivariate effective sample size and the stopping rules of \citet{jone:hara:caff:neat:2006} and \citet{vats:fleg:jone:2019}.

Geometric ergodicity of block Gibbs samplers for Bayesian linear mixed models with \emph{normal} random effects is well understood \citep{hobe:geye:1998,jone:hobe:2004,john:jone:2010,roma:hobe:2012,roma:hobe:2014,abra:hobe:2017}, and related results exist for generalized linear mixed models \citep{wang:roy:2018a,wang:roy:2018b,rao:roy:2021}. A parallel literature studies regression models whose \emph{errors} are scale mixtures of normals, where conditions on the mixing density are given for data augmentation algorithms \citep{RoyHobert2010,hobe:jung:khar:qin:2018,qin:hobe:2018,back:hobe:2020,hamu:2024}.

Global-local shrinkage priors also have the structure of a scale mixture of normals, and the convergence of the associated Gibbs samplers has received much attention \citep{KhareHobert2013,PalKhare2014,JohndrowOrensteinBhattacharya2020,bhat:khar:pal:2022,hamu:2025}. The random-effects distribution considered here has the same form as a global-local prior on the random effects, with the random-effects precision as the global and the latent weights as the local precisions, but these results do not cover our setting. In the shrinkage literature the prior is scaled by the error variance, which allows the error variance and the coefficients to be updated jointly, and there is no separate block of fixed effects with a proper prior; \citet{KhareHobert2013} and \citet{PalKhare2014}, for instance, treat arbitrary designs under that scaling. \citet{abra:hobe:2019} consider a linear mixed model with a normal-gamma prior on the fixed effects and normal random effects, and analyse a hybrid algorithm that is easier to study than the three-block Gibbs sampler. Heavy-tailed hierarchical models have also been studied from the point of view of parametrization: \citet{papa:robe:2008} show that the stability of the Gibbs sampler that updates the random effects and their location directly depends on the relative tail behaviour of the levels of the hierarchy. The sampler studied here uses the latent weights instead, and in our setting, with normal errors, the conditions involve only the mixing distribution, the random-effects design and the priors of the variance components.

The closest work is the recent preprint of \citet{awad:boli:jin:sima:wall:2026}, which studies latent Gaussian fields driven by generalized inverse Gaussian variance mixtures. There the regression and variance parameters are held fixed, the sampler alternates between the latent field and the mixing variables, and the analysis combines a trace-class argument with a drift condition that involves the null space of the observation operator. In the model considered here the fixed effects and both precisions are random and updated within the three-block sampler, the mixing distribution is arbitrary subject to conditions near the origin, and for the usual families no condition on the null space of the design is needed. To our knowledge, there are no convergence results for the data augmentation Gibbs sampler of a linear mixed model with scale-mixture random effects and random variance components.

Our contributions are the following.
\begin{enumerate}[leftmargin=*]
\item A general criterion for geometric ergodicity of the three-block Gibbs sampler, for an arbitrary mixing distribution $H$ and an arbitrary random-effects design matrix $Z$ (Theorem~\ref{thm:general}). It requires a positive moment of $H$ of some order larger than $1/2$ and controls the growth of a negative moment of the exponentially tilted laws $H_s(d\eta)\propto\eta^{1/2}e^{-s\eta}H(d\eta)$, which are the full conditionals of the weights. The proof uses a drift function built from fractional powers, which is what allows such weak moment conditions; when $Z$ has full column rank, the criterion covers the Student-$t$ law with any degrees of freedom, including $d\le1$, for which the random effects have no mean.
\item A rescaling argument that removes the dependence on the design. The Gibbs sampler is equivariant under a joint rescaling of the columns of $Z$ and of the mixing distributions, and a suitable rescaling concentrates the null space of $Z$ on a few coordinates. As a result, the Student-$t$ and slash families with any tail parameter, the contaminated normal, the normal-generalized inverse Gaussian mixtures and the variance-gamma laws with shape larger than $1/2$, including the Laplace law, are covered for every design, full rank or not, with the rank entering only the conditions on the priors of the variance components (Theorem~\ref{thm:anydesign} and Corollary~\ref{cor:families}). The result for full-rank $Z$ also extends to random effects whose scale depends on a known grouping (Proposition~\ref{prop:groups}).
\item Consequences for Monte Carlo error assessment. Posterior moment bounds that hold for every mixing distribution, combined with geometric ergodicity, give a Markov chain CLT and consistent multivariate batch-means estimators for the regression coefficients and the log-variance components, and hence Monte Carlo standard errors and effective-sample-size stopping rules (Section~\ref{sec:practice}). Three simulation studies examine misspecification of the random-effects distribution, the coverage of Monte Carlo confidence regions at practical run lengths, and a rank-deficient design with crossed random factors. In the analysis of longitudinal $\beta$-HCG data from early pregnancies \citep{MarshallBaron2000}, the latent weights of the heavy-tailed models point to larger between-subject variability among abnormal pregnancies, and a model with group-specific scales fits better than any model with a common scale.
\end{enumerate}
The model class has one latent weight per random effect, a single scale (or scales indexed by a known grouping) and normal errors. Multivariate scale-mixture random effects, in which the coordinates of a subject's random-effect vector share one weight, and heavy-tailed errors are not covered (Section~\ref{sec:disc}). Code that implements the sampler for all the families above, together with the output analysis, is publicly available (see the section on code and data at the end of the paper).

The paper is organized as follows. Section~\ref{sec:model} describes the model and the Gibbs sampler. Section~\ref{sec:results} states the main results and applies them to specific families. Section~\ref{sec:proofs} contains the proofs. Section~\ref{sec:practice} discusses the practical implications of geometric ergodicity. Sections~\ref{sec:sim} and~\ref{sec:app} present the simulation studies and the application, and Section~\ref{sec:disc} concludes. Appendix~\ref{app:KH} states an auxiliary matrix result, Appendix~\ref{app:code} describes how the implementation was verified, and Appendix~\ref{app:covdiag} reports diagnostics for the coverage study.

%% file: sec2_model.tex
\section{Model and block Gibbs sampler}\label{sec:model}

\subsection{The model}
Let $Y$ be an $N\times1$ response vector, $X$ an $N\times p$ fixed-effects design matrix and $Z$ an $N\times q$ random-effects design matrix, both known. Let $H$ be a probability distribution on $\R_+=(0,\infty)$. We consider the hierarchical model
\begin{subequations}\label{eq:model}
\begin{align}
Y\mid\beta,u,\lambda_e &\sim \N_N\big(X\beta+Zu,\ \lambda_e^{-1}I_N\big),\label{eq:model-a}\\
u_i\mid\lambda_u,\eta &\ind \N\big(0,(\lambda_u\eta_i)^{-1}\big),\qquad \eta_1,\dots,\eta_q\iid H,\label{eq:model-b}\\
\beta\sim\N_p(\mu_\beta,\Sigma_\beta),\quad &\lambda_e\sim\Ga(a_e,b_e),\quad \lambda_u\sim\Ga(a_u,b_u),\label{eq:model-c}
\end{align}
\end{subequations}
with $\beta$, $\lambda_e$, $\lambda_u$ and $\eta=(\eta_1,\dots,\eta_q)^T$ independent a priori. Here $\Ga(a,b)$ denotes the gamma distribution with density proportional to $x^{a-1}e^{-bx}$; the hyperparameters $\mu_\beta$, $\Sigma_\beta$ (positive definite) and $a_e,b_e,a_u,b_u>0$ are known. Integrating out $\eta_i$, the random effect $u_i$ given $\lambda_u$ has density
\[
f(u\mid\lambda_u)=\int_0^\infty \Big(\frac{\lambda_u\eta}{2\pi}\Big)^{1/2}\exp\Big(-\frac{\lambda_u\eta u^2}{2}\Big)H(d\eta),
\]
a scale mixture of normals with scale parameter $\lambda_u^{-1/2}$ \citep{Andrews1974,West1987}. The posterior distribution is proper for every $H$: the likelihood is bounded by $(\lambda_e/2\pi)^{N/2}$, which is integrable against the gamma prior of $\lambda_e$, and all other priors are proper. Table~\ref{tab:families} lists the families used in this paper. We parametrize every family through the \emph{precision} multiplier $\eta$; small values of $\eta_i$ inflate the variance of $u_i$.

\begin{table}[t]
\centering\small
\caption{Scale mixtures of normals used in this paper, the mixing distribution $H$ of the precision multiplier $\eta$, and the full conditional $H_s$ of $\eta_i$ in \eqref{eq:tilted}. $\GIG(\varrho,\chi,\psi)$ has density proportional to $x^{\varrho-1}\exp\{-(\chi/x+\psi x)/2\}$; $\mathrm{IG}(\alpha,\beta)$ is the inverse gamma distribution.}
\label{tab:families}
\renewcommand{\arraystretch}{1.2}
\begin{tabular}{>{\raggedright\arraybackslash}p{3.4cm}>{\raggedright\arraybackslash}p{4.1cm}>{\raggedright\arraybackslash}p{6.2cm}}
\toprule
Distribution of $u_i\mid\lambda_u$ & Mixing distribution $H$ & Full conditional $H_s$, $s=\lambda_u u_i^2/2$\\
\midrule
Normal & point mass at 1 & point mass at 1\\
Student-$t$, $d>0$ d.f. & $\Ga(d/2,d/2)$ & $\Ga\big((d+1)/2,\,d/2+s\big)$\\
Slash, $\nu>0$ & $\mathrm{Beta}(\nu,1)$ & $\Ga(\nu+1/2,s)$ truncated to $(0,1)$ if $s>0$; $\mathrm{Beta}(\nu+1/2,1)$ if $s=0$\\
Contaminated normal, $\nu,\gamma\in(0,1)$ & $\nu\,\delta_\gamma+(1-\nu)\,\delta_1$ & two-point law on $\{\gamma,1\}$ with odds $\nu\gamma^{1/2}e^{-s\gamma}:(1-\nu)e^{-s}$\\
Normal-GIG (NIG, hyperbolic) & $\GIG(\varrho,\chi,\psi)$, $\chi>0$, $\psi>0$ & $\GIG(\varrho+1/2,\chi,\psi+2s)$\\
Variance-gamma, shape $\alpha$ & $\mathrm{IG}(\alpha,\beta)=\GIG(-\alpha,2\beta,0)$ & $\GIG(1/2-\alpha,2\beta,2s)$\\
Laplace & $\mathrm{IG}(1,1/2)$ & inverse Gaussian, mean $(2s)^{-1/2}$, shape 1, if $s>0$; $\mathrm{IG}(1/2,1/2)$ if $s=0$\\
\bottomrule
\end{tabular}
\end{table}

\subsection{Full conditionals and the sampler}
Write $\lambda=(\lambda_e,\lambda_u)$, $\theta=(\beta^T,u^T)^T$, $W=(X\ Z)$ and $\Deta=\diag(\eta_1,\dots,\eta_q)$. The joint posterior density of $(\lambda,\eta,\theta)$ with respect to $d\lambda\,H^{\otimes q}(d\eta)\,d\theta$ is proportional to
\[
\lambda_e^{N/2}e^{-\lambda_e\norm{y-W\theta}^2/2}\,e^{-(\beta-\mu_\beta)^T\Sigma_\beta^{-1}(\beta-\mu_\beta)/2}\,\lambda_u^{q/2}\prod_{i=1}^q\eta_i^{1/2}e^{-\lambda_u\eta_iu_i^2/2}\,\lambda_e^{a_e-1}e^{-b_e\lambda_e}\lambda_u^{a_u-1}e^{-b_u\lambda_u}.
\]
Collecting, for each block, the factors of this density that involve it gives the following full conditionals.
\begin{itemize}[leftmargin=*]
\item Given $(\eta,\theta)$, $\lambda_e$ and $\lambda_u$ are independent with
$\lambda_e\sim\Ga\big(a_e+N/2,\ b_e+\norm{y-W\theta}^2/2\big)$ and $\lambda_u\sim\Ga\big(a_u+q/2,\ b_u+\norm{\Deta^{1/2}u}^2/2\big)$.
\item Given $(\lambda,\theta)$, the $\eta_i$ are independent and $\eta_i\sim H_{s_i}$ with $s_i=\lambda_u u_i^2/2$, where for $s\ge0$
\begin{equation}\label{eq:tilted}
H_s(d\eta)=\frac{\eta^{1/2}e^{-s\eta}H(d\eta)}{\int_0^\infty t^{1/2}e^{-st}H(dt)}.
\end{equation}
For $s>0$ the denominator is finite and positive; for $s=0$ it is $m_{1/2}=\int\eta^{1/2}H(d\eta)$, which we assume finite in Sections~\ref{sec:results} and~\ref{sec:proofs}.
\item Given $(\lambda,\eta)$, $\theta$ is normal with precision matrix and mean
\[
P_{\lambda,\eta}=\begin{pmatrix}T_\lambda & \lambda_eX^TZ\\ \lambda_eZ^TX & \lambda_eZ^TZ+\lambda_u\Deta\end{pmatrix},\qquad
\E(\theta\mid\lambda,\eta)=P_{\lambda,\eta}^{-1}\begin{pmatrix}\lambda_eX^Ty+\Sigma_\beta^{-1}\mu_\beta\\ \lambda_eZ^Ty\end{pmatrix},
\]
where $T_\lambda=\lambda_eX^TX+\Sigma_\beta^{-1}$.
\end{itemize}
Let $M_\lambda=I_N-\lambda_eXT_\lambda^{-1}X^T$ and $Q_{\lambda,\eta}=\lambda_eZ^TM_\lambda Z+\lambda_u\Deta$. Taking the Schur complement of $T_\lambda$ in $P_{\lambda,\eta}$ gives
\begin{equation}\label{eq:umoments}
\Var(u\mid\lambda,\eta)=Q_{\lambda,\eta}^{-1},\qquad
\E(u\mid\lambda,\eta)=\lambda_eQ_{\lambda,\eta}^{-1}Z^T\big(M_\lambda y-XT_\lambda^{-1}\Sigma_\beta^{-1}\mu_\beta\big).
\end{equation}
The expression for $\Var(u\mid\lambda,\eta)$ is used in Section~\ref{sec:proofs}.

The block Gibbs sampler is described in Algorithm~\ref{alg:gibbs}. It simulates a Markov chain $\Phi=\{(\lambda^{(n)},\eta^{(n)},\theta^{(n)})\}_{n\ge0}$ on $\mathsf X_\Phi=\R_+^2\times\mathsf S^q\times\R^{p+q}$, where $\mathsf S\subseteq\R_+$ is the support of $H$, and the posterior is its invariant distribution. The order of the updates matters for the analysis: since $\lambda$ is updated first from a distribution that depends only on $(\eta,\theta)$, the sequence $\{(\eta^{(n)},\theta^{(n)})\}$ is itself a Markov chain, which we call the \emph{marginal chain}.

\begin{algorithm}[t]
\caption{Block Gibbs sampler for model \eqref{eq:model}}\label{alg:gibbs}
\begin{algorithmic}[1]
\Require current state $(\eta,\theta)$ with $\theta=(\beta,u)$
\State Draw $\lambda_e\sim\Ga\big(a_e+N/2,\,b_e+\norm{y-X\beta-Zu}^2/2\big)$ and $\lambda_u\sim\Ga\big(a_u+q/2,\,b_u+\sum_i\eta_iu_i^2/2\big)$.
\State For $i=1,\dots,q$, draw $\eta_i\sim H_{s_i}$ with $s_i=\lambda_uu_i^2/2$ (Table~\ref{tab:families}).
\State Draw $\theta\sim\N_{p+q}\big(\E(\theta\mid\lambda,\eta),P_{\lambda,\eta}^{-1}\big)$.
\end{algorithmic}
\end{algorithm}

For the families in Table~\ref{tab:families} the second step is a draw from a standard distribution. When $Z^TZ$ is diagonal, as in random-intercept models, the third step can be carried out exactly at a cost that is linear in $q$ once $X^TX$, $X^TZ$ and $Z^Ty$ have been precomputed: one draws $\beta$ from its marginal normal distribution given $(\lambda,\eta)$, whose precision is the Schur complement $T_\lambda-\lambda_e^2X^TZ(\lambda_eZ^TZ+\lambda_u\Deta)^{-1}Z^TX$, and then the $u_i$ independently given $\beta$.

%% file: sec3_results.tex
\section{Main results}\label{sec:results}

\subsection{Quantities associated with the mixing distribution and the design}
For $r\in\R$ let $m_r=\int_0^\infty\eta^rH(d\eta)\in(0,\infty]$, and write $\E_s$ for expectation under the tilted law $H_s$ in \eqref{eq:tilted}. For $c\in(0,1]$ let
\[
M_c=\sup_{s\ge0}\E_s[\eta^c],\qquad
\kappa_H(c)=\limsup_{s\to\infty}\frac{\E_s[\eta^{-c}]}{s^c},\qquad
R_H(c)=\frac{\Gamma(c+\frac12)}{\Gamma(\frac12)}\,\kappa_H(c).
\]
By Lemma~\ref{lem:tilted} below, $M_c=m_{c+1/2}/m_{1/2}$, so $M_c<\infty$ if and only if $m_{c+1/2}<\infty$. The quantity $\kappa_H(c)$ measures how fast the conditional mean of $\eta_i^{-c}$ grows with $u_i^2$. Since $\eta_i^{-1}$ is the conditional variance multiplier of $u_i$, $\kappa_H(c)$ is governed by the behaviour of $H$ near zero, that is, by the tails of the marginal distribution of $u_i$. The factor $\Gamma(c+\frac12)/\Gamma(\frac12)=2^{-c}\E|\xi|^{2c}$, with $\xi\sim\N(0,1)$, comes from the fractional moments of a normal variable. For the Student-$t$ distribution with $d$ degrees of freedom, $H_s=\Ga(\frac{d+1}2,\frac d2+s)$, so
\[
\E_s[\eta^{-c}]=\frac{\Gamma(\frac{d+1}2-c)}{\Gamma(\frac{d+1}2)}\Big(\frac d2+s\Big)^c,\qquad
R_H(c)=\frac{\Gamma(c+\frac12)\,\Gamma(\frac{d+1}2-c)}{\Gamma(\frac12)\,\Gamma(\frac{d+1}2)}\qquad\Big(c<\frac{d+1}2\Big).
\]
In particular, for $d>1$, $R_H(1)=1/(d-1)$.

For the random-effects design matrix let $r=\rank(Z)$, let $\Pi_Z$ denote the orthogonal projection of $\R^q$ onto the null space $\mathcal N(Z)=\{v\in\R^q:Zv=0\}$, and let
\[
\ell_i=(\Pi_Z)_{ii},\qquad \ell^*=\max_{1\le i\le q}\ell_i,\qquad S_Z(c)=\max_{1\le j\le q}\sum_{i=1}^q|(\Pi_Z)_{ij}|^{2c}\quad(c\in(0,1]).
\]
Thus $0\le\ell_i\le1$ and $\sum_i\ell_i=q-r$. Since $\sum_i(\Pi_Z)_{ij}^2=(\Pi_Z)_{jj}$, $S_Z(1)=\ell^*$. Moreover, $S_Z(c)=0$ exactly when $Z$ has full column rank, and $S_Z(c)=1$ for every $c$ when the rank deficiency comes only from zero columns of $Z$, because $\Pi_Z$ is then a diagonal matrix of zeros and ones. The quantity $S_Z(c)$ can be computed from a basis of $\mathcal N(Z)$. We also write $\tau_Z=\tr\{(Z^TZ)^+\}$, where $A^+$ is the Moore--Penrose inverse.

\subsection{Geometric ergodicity}
Recall that a Harris ergodic Markov chain with transition kernel $P$ and invariant distribution $\pi$ is geometrically ergodic if there exist $\rho<1$ and a function $C$ with $\norm{P^n(x,\cdot)-\pi}_{\mathrm{TV}}\le C(x)\rho^n$ for all $x$ and $n$.

\begin{theorem}\label{thm:general}
Suppose that
\begin{enumerate}[label=\textup{(C\arabic*)},leftmargin=*]
\item\label{C1} $a_e>\frac12(r-N+2)$;
\item\label{C2} if $r<q$, $a_u>1-\frac r2$.
\end{enumerate}
Suppose also that there is $c\in(0,1]$ such that the mixing distribution satisfies
\begin{enumerate}[label=\textup{(H\arabic*)},leftmargin=*]
\item\label{H1} $m_{c+1/2}<\infty$ and $\kappa_H(c)<\infty$;
\item\label{H2} $R_H(c)\,S_Z(c)<1$.
\end{enumerate}
Then the block Gibbs chain $\Phi$ is geometrically ergodic.
\end{theorem}

When $Z$ has full column rank, \ref{C2} and \ref{H2} are void, and the theorem requires only \ref{C1} and \ref{H1} for some $c\in(0,1]$. When $Z$ is rank deficient, \ref{H2} is the only condition of Theorem~\ref{thm:general} that couples the mixing distribution with the design. The choice of $c$ involves a trade-off. Taking $c=1$ gives the condition $\ell^*R_H(1)<1$, which for Student-$t$ random effects reads $d>1+\ell^*$. Smaller values of $c$ weaken the moment condition in \ref{H1}, and for families with an index at zero, a notion defined in Section~\ref{sec:verify} (Proposition~\ref{prop:families}(c)), $R_H(c)\to1$ as $c\downarrow0$, but $S_Z(c)$ does not decrease as $c$ decreases.

Theorem~\ref{thm:general} depends on how the columns of $Z$ are scaled, through $S_Z(c)$, whereas the Gibbs sampler is equivariant under a joint rescaling of the columns of $Z$ and of the mixing distributions. Exploiting this gives a condition that does not involve the design at all.

\begin{theorem}\label{thm:anydesign}
Suppose that \ref{C1}--\ref{C2} hold and that there is $c\in(0,1]$ such that \ref{H1} holds and, if $r<q$, $R_H(c)<1$. Then the block Gibbs chain $\Phi$ is geometrically ergodic.
\end{theorem}

The proof (Section~\ref{sec:proof-anydesign}) multiplies $q-r$ suitably chosen coordinates of $u$ by a large constant. In the rescaled but equivalent model, the null space of the design concentrates on those coordinates, which then behave like zero columns of $Z$, and Theorem~\ref{thm:general}, extended to coordinate-specific mixing distributions, applies. For the families with an index at zero, $R_H(c)<1$ for all small $c>0$ (see the proof of Corollary~\ref{cor:families}(i), which uses only the index), and for families with $\kappa_H(c)=0$ the condition is void. Theorem~\ref{thm:general} also shows how the projection onto the null space enters the drift, and it covers mixing distributions with $R_H(c)\ge1$ for all $c$ when $S_Z$ is small.

Theorem~\ref{thm:general} extends to random effects whose scale depends on a known grouping of the subjects, for example a treatment arm or a diagnostic group. Let $k:\{1,\dots,q\}\to\{1,\dots,G\}$ be a known labelling and replace \eqref{eq:model-b}--\eqref{eq:model-c} by
\begin{equation}\label{eq:groupscale}
u_i\mid\lambda,\eta\ind\N\big(0,(\lambda_{u,k(i)}\eta_i)^{-1}\big),\qquad\lambda_{u,1},\dots,\lambda_{u,G}\ \text{independent},\ \lambda_{u,k}\sim\Ga(a_{u,k},b_{u,k}).
\end{equation}
The Gibbs sampler is the same, with $\lambda=(\lambda_e,\lambda_{u,1},\dots,\lambda_{u,G})$ updated from independent gamma full conditionals and $s_i=\lambda_{u,k(i)}u_i^2/2$ in the second step.

\begin{proposition}\label{prop:groups}
Under model \eqref{eq:groupscale}, suppose that $\rank(Z)=q$, that \ref{C1} holds, and that \ref{H1} holds for some $c\in(0,1]$. Then the block Gibbs chain is geometrically ergodic.
\end{proposition}

The case of several scales with a rank-deficient $Z$, which includes random intercepts and slopes with separate scales, requires a block version of the argument (Section~\ref{sec:disc}).

\begin{remark}\label{rem:normalcase}
For normal random effects ($H$ a point mass at one), $\kappa_H(c)=0$ and \ref{H1}--\ref{H2} hold for every $c$. Conditions \ref{C1}--\ref{C2} then have the same form as the conditions of \citet{roma:hobe:2014} for the block Gibbs sampler with proper priors, without their requirement that $X$ have full column rank. By Theorem~\ref{thm:anydesign}, heavy-tailed random effects require no condition beyond \ref{H1} and $R_H(c)<1$, which holds for all the families in Table~\ref{tab:families} (for the variance-gamma family, when its shape exceeds $\frac12$; Corollary~\ref{cor:families}).
\end{remark}

\begin{remark}\label{rem:leverage}
The null-space projection enters through Lemma~\ref{lem:trace}(a), which bounds the conditional variance of each $u_i$ along $\mathcal N(Z)$ by a combination of the $\eta_j^{-1}$ with weights $(\Pi_Z)_{ij}^2$. If the rank deficiency is instead bounded by its dimension $q-r$, as in the trace inequality of \citet[Lemma~3]{roma:hobe:2014}, the same argument with $c=1$ gives $(q-r)R_H(1)<1$, which for Student-$t$ random effects reads $d>1+q-r$.

As an example, suppose $Z=(Z_1\ Z_2)$ contains the indicator matrices of two crossed factors with $q_1$ and $q_2$ levels, so that the columns of $Z_1$ and of $Z_2$ both sum to $1_N$. If $Z$ has no other linear dependencies, $\mathcal N(Z)$ is spanned by $(1_{q_1}^T,-1_{q_2}^T)^T$. Then every entry of $\Pi_Z$ has absolute value $1/(q_1+q_2)$, and $S_Z(c)=(q_1+q_2)^{1-2c}$. Theorem~\ref{thm:general} alone gives the following.
\begin{itemize}[leftmargin=*]
\item With $c=1$, \ref{H2} becomes $R_H(1)<q_1+q_2$, that is, $d>1+1/(q_1+q_2)$ for Student-$t$ random effects.
\item With $c=\frac12$, $S_Z=1$, and \ref{H2} holds for every $d>1$.
\item Minimizing over $c$ gives smaller thresholds still: for $q_1=q_2=5$, every $d\ge0.75$.
\end{itemize}
Theorem~\ref{thm:anydesign} shows that in this example every $d>0$ is in fact allowed. When the rank deficiency comes only from zero columns, for instance subjects included for prediction without observations, $S_Z\equiv1$ and \ref{H2} reduces to the condition $R_H(c)<1$ of Theorem~\ref{thm:anydesign}.
\end{remark}

\subsection{Verifying the conditions}\label{sec:verify}
The next proposition computes $\kappa_H(c)$ for three types of mixing distributions. Following \citet{hobe:jung:khar:qin:2018}, we distinguish mixing distributions that vanish near zero, have a GIG form, or behave like a power near zero. We say that $H$ has \emph{index $a>0$ at zero} if, on some interval $(0,\epsilon)$, $H$ has a density of the form
\begin{equation}\label{eq:regvar}
h(\eta)=\eta^{a-1}L(\eta),
\end{equation}
where $L$ is slowly varying at zero, that is, $L(tx)/L(x)\to1$ as $x\downarrow0$ for every $t>0$.

\begin{proposition}\label{prop:families}
Assume $m_{1/2}<\infty$ and let $c\in(0,1]$.
\begin{enumerate}[label=\textup{(\alph*)},leftmargin=*]
\item If $H((0,\delta))=0$ for some $\delta>0$, then $\kappa_H(c)=0$.
\item If $H=\GIG(\varrho,\chi,\psi)$ with $\chi>0$, and either $\psi>0$, or $\psi=0$ and $\varrho<-\frac12$, then $\kappa_H(c)=0$.
\item If $H$ has index $a$ at zero and $c<a+\frac12$, then $\kappa_H(c)=\Gamma(a+\frac12-c)/\Gamma(a+\frac12)$, and the $\limsup$ in its definition is a limit.
\end{enumerate}
\end{proposition}

Combining Theorems~\ref{thm:general} and~\ref{thm:anydesign} with Proposition~\ref{prop:families} gives the following conclusions. Conditions \ref{C1}--\ref{C2} are assumed throughout.

\begin{corollary}\label{cor:families}
\begin{enumerate}[label=\textup{(\roman*)},leftmargin=*]
\item \emph{Student-$t$, $d>0$}: $H=\Ga(d/2,d/2)$ has all positive moments and index $a=d/2$. The chain is geometrically ergodic for every $d>0$ and every design $Z$. With $c=1$, Theorem~\ref{thm:general} alone gives the explicit condition $d>1+\ell^*$.
\item \emph{Slash, $\nu>0$}: $H=\mathrm{Beta}(\nu,1)$ has bounded support and index $a=\nu$. The chain is geometrically ergodic for every $\nu>0$ and every design; with $c=1$, which is allowed when $\nu>\frac12$, $R_H(1)=\Gamma(\frac32)\Gamma(\nu-\frac12)/\{\Gamma(\frac12)\Gamma(\nu+\frac12)\}=1/(2\nu-1)$, and Theorem~\ref{thm:general} gives $\nu>(1+\ell^*)/2$.
\item \emph{Contaminated normal}: by Proposition~\ref{prop:families}(a) with $\delta=\gamma$, the chain is geometrically ergodic for every $Z$.
\item \emph{Normal-GIG with $\chi,\psi>0$}: all moments of $H$ are finite, and the chain is geometrically ergodic for every $Z$. Because $\eta$ is a precision, the normal inverse Gaussian law corresponds to $\eta\sim\GIG(\frac12,\psi,\chi)$ and the hyperbolic law to $\eta\sim\GIG(-1,\psi,\chi)$, where $(\chi,\psi)$ are the parameters of the GIG law of the variance $1/\eta$; only their symmetric versions are covered.
\item \emph{Variance-gamma with shape $\alpha$}: $m_r<\infty$ if and only if $r<\alpha$. The chain is geometrically ergodic for every $Z$ if $\alpha>1/2$. In particular, Laplace random effects ($\alpha=1$) are covered for every design.
\end{enumerate}
\end{corollary}

\begin{remark}\label{rem:boundary}
If $\alpha\le1/2$ in Corollary~\ref{cor:families}(v), then $m_{1/2}=\infty$, the marginal density of $u_i$ is unbounded at the origin, and the law $H_0$ in \eqref{eq:tilted} is not defined. The posterior is still proper and the sampler is still well defined, because $s_i>0$ almost surely. This case, however, falls outside the present framework, which uses $m_{1/2}<\infty$ throughout, and we do not know whether the chain is geometrically ergodic there. The horseshoe-type mixing density $h(\eta)\propto\eta^{-1/2}(1+\eta)^{-1}$ is in the same situation. Such distributions are used as shrinkage priors rather than as models for random effects, and their Gibbs samplers have been studied with different arguments \citep{JohndrowOrensteinBhattacharya2020,bhat:khar:pal:2022}.
\end{remark}

%% file: sec4_proofs.tex
\section{Proofs}\label{sec:proofs}

For symmetric matrices $A$ and $B$ we write $A\pe B$ if $B-A$ is non-negative definite. We use repeatedly that $A\pe B$ implies $\tr(A)\le\tr(B)$ and $C^TAC\pe C^TBC$ for any conformable $C$, and that for positive definite $A,B$, $A\pe B$ if and only if $B^{-1}\pe A^{-1}$. Norms of vectors are Euclidean and $\norm{A}_2$ is the spectral norm of a matrix.

\subsection{Reduction to the marginal chain and topological properties}

Let $P_m$ denote the Markov transition kernel of the marginal chain $\{(\eta^{(n)},\theta^{(n)})\}$ on $\mathsf X=\mathsf S^q\times\R^{p+q}$, and let $\pi_m$ be the $(\eta,\theta)$-marginal of the posterior. Starting from $(\eta',\theta')$, one step draws $\lambda\sim\pi(\lambda\mid\eta',\theta')$, then $\eta\sim\pi(\eta\mid\lambda,\theta')$, then $\theta\sim\pi(\theta\mid\lambda,\eta)$. Each of these draws leaves the posterior invariant, so $\pi_m$ is invariant for $P_m$. Moreover $P_m$ has a density with respect to $\varphi=H^{\otimes q}\otimes\mathrm{Leb}_{p+q}$,
\begin{equation}\label{eq:kdens}
k(\eta,\theta\mid\eta',\theta')=\int_{\R_+^2}\pi(\theta\mid\lambda,\eta)\prod_{i=1}^q\frac{\eta_i^{1/2}e^{-\lambda_u u_i'^2\eta_i/2}}{\int t^{1/2}e^{-\lambda_uu_i'^2t/2}H(dt)}\,\pi(\lambda\mid\eta',\theta')\,d\lambda ,
\end{equation}
which is strictly positive on $\mathsf X\times\mathsf X$. We write $P_\Phi$ for the transition kernel of $\Phi$.

\begin{lemma}\label{lem:topology}
Assume $m_{1/2}<\infty$. The marginal chain and the Gibbs chain $\Phi$ are Harris ergodic, and every compact subset of $\mathsf X$ is petite for the marginal chain. Moreover, if the marginal chain is geometrically ergodic, so is $\Phi$.
\end{lemma}

\begin{proof}
Since $k>0$, $P_m((\eta',\theta'),A)>0$ for every $(\eta',\theta')$ whenever $\varphi(A)>0$; hence the chain is $\varphi$-irreducible. It is also aperiodic: if $D_1,\dots,D_{\varpi}$ with $\varpi\ge2$ were a periodic cycle, then $\varphi(D_1)>0$, because $P_m(x,D_1)=1$ for $x\in D_\varpi$ and $P_m(x,\cdot)$ has a density with respect to $\varphi$; and $P_m(x,D_1)=0$ for $x\in D_1$, which contradicts the previous sentence. The posterior density is strictly positive, so $\pi_m$ is equivalent to $\varphi$; since $P_m((\eta',\theta'),\cdot)\ll\varphi$, we get $P_m((\eta',\theta'),\cdot)\ll\pi_m$ for every starting point, and Harris recurrence follows from Corollary~1 of \citet{tier:1994}. With the invariant probability $\pi_m$, the chain is positive Harris recurrent and aperiodic, that is, Harris ergodic. The same argument applies to $\Phi$: its kernel has the strictly positive density $\pi(\theta\mid\lambda,\eta)\,\pi(\eta\mid\lambda,\theta')\,\pi(\lambda\mid\eta',\theta')$ with respect to $\mathrm{Leb}_2\otimes H^{\otimes q}\otimes\mathrm{Leb}_{p+q}$, to which the posterior $\pi$ is equivalent, and $\pi$ is invariant for $P_\Phi$. The integrand in \eqref{eq:kdens} is continuous in $(\eta',\theta')$, and it is dominated, locally uniformly in $(\eta',\theta')$, by an integrable function of $\lambda$: the normal density $\pi(\theta\mid\lambda,\eta)$ is bounded by a polynomial in $(\lambda_e,\lambda_u)$, and the gamma densities $\pi(\lambda\mid\eta',\theta')$ have rate parameters bounded below by $b_e$ and $b_u$ and above on compact sets. For the $i$th tilted factor, write $s_i=\lambda_uu_i'^2/2$; since $\eta_i\in\mathsf S$, for every $\delta>0$ the denominator is at least $e^{-s_i(\eta_i+\delta)}\int_{(0,\eta_i+\delta)}t^{1/2}H(dt)>0$, so the factor is at most $C_{\delta}(\eta_i)e^{s_i\delta}$. If $|u_i'|\le\bar u$ for all $i$ on a neighbourhood of $(\eta',\theta')$ and $q\delta\bar u^2<b_u$, the product of the $q$ factors is at most a constant times $e^{\lambda_ub_u/2}$, which is integrable against the gamma density of $\lambda_u$, whose rate is at least $b_u$. By dominated convergence $k$ is continuous in $(\eta',\theta')$, and by Scheff\'e's lemma $(\eta',\theta')\mapsto P_m((\eta',\theta'),\cdot)$ is continuous in total variation; thus the chain is (strong, hence weak) Feller. The space $\mathsf X$ is a closed subset of $\R_+^q\times\R^{p+q}$, hence locally compact and separable, and $\mathrm{supp}(\varphi)=\mathsf X$ has non-empty interior in $\mathsf X$. By Theorems~6.2.9 and~6.2.5 of \citet{meyn:twee:1993}, the chain is a T-chain and every compact set is petite.

For the last statement, note that $\Phi^{(n+1)}=(\lambda^{(n+1)},\eta^{(n+1)},\theta^{(n+1)})$ is obtained from $(\eta^{(n)},\theta^{(n)})$ by applying the fixed kernel $R$ that draws $\lambda$, $\eta$ and $\theta$ in turn, and that $\pi_mR=\pi$. Therefore, for any initial state $x=(\lambda^{(0)},\eta^{(0)},\theta^{(0)})$ and $n\ge0$,
\[
\norm{P_\Phi^{n+1}(x,\cdot)-\pi}_{\mathrm{TV}}\le\norm{P_m^n((\eta^{(0)},\theta^{(0)}),\cdot)-\pi_m}_{\mathrm{TV}},
\]
because applying a Markov kernel does not increase total variation distance. If the right-hand side is at most $C_m(\eta^{(0)},\theta^{(0)})\rho^n$, then $\norm{P_\Phi^{n}(x,\cdot)-\pi}_{\mathrm{TV}}\le\max\{1,C_m(\eta^{(0)},\theta^{(0)})/\rho\}\rho^n$ for all $n\ge0$.
\end{proof}

By Lemma~\ref{lem:topology} and Lemma~15.2.8 and Theorem~15.0.1 of \citet{meyn:twee:1993}, Theorem~\ref{thm:general} and Proposition~\ref{prop:groups}, and through Step~1 of its proof also Theorem~\ref{thm:anydesign}, follow once we exhibit a function $v:\mathsf X\to[1,\infty)$ whose sublevel sets are compact and which satisfies the drift condition
\begin{equation}\label{eq:drift}
\E\big[v(\eta,\theta)\mid\eta',\theta'\big]\le\rho\,v(\eta',\theta')+L\qquad\text{for all }(\eta',\theta')\in\mathsf X,
\end{equation}
for some $\rho<1$ and $L<\infty$, where the expectation is taken with respect to $P_m((\eta',\theta'),\cdot)$.

\subsection{Matrix inequalities}

The following lemma gives coordinatewise bounds on the inverse of a matrix of the form $\upsilon A^TA+\Upsilon$, separating the row space of $A$ from its null space. Part (b) refines Lemma~3 of \citet{roma:hobe:2014}.

\begin{lemma}\label{lem:trace}
Let $A$ be an $m\times n$ matrix, $\upsilon>0$, and $\Upsilon=\diag(\upsilon_1,\dots,\upsilon_n)$ with $\upsilon_i>0$. Let $\Omega=\upsilon A^TA+\Upsilon$, let $\Pi_A$ be the orthogonal projection of $\R^n$ onto $\mathcal N(A)$, $P_A=I-\Pi_A$, and $\ell_i=(\Pi_A)_{ii}$. Then
\begin{enumerate}[label=\textup{(\alph*)},leftmargin=*]
\item $P_A\Omega^{-1}P_A\pe\upsilon^{-1}(A^TA)^+$ and $\Pi_A\Omega^{-1}\Pi_A\pe\Pi_A\Upsilon^{-1}\Pi_A$; the same bounds hold with $\Omega^{-1}$ replaced by $Q^{-1}$ for any $Q\succeq\Omega$;
\item $\tr(\Omega^{-1})\le\upsilon^{-1}\tr\{(A^TA)^+\}+\sum_{i=1}^n\ell_i\,\upsilon_i^{-1}$;
\item $\tr(A\Omega^{-1}A^T)\le\upsilon^{-1}\rank(A)$.
\end{enumerate}
\end{lemma}

\begin{proof}
(a) Since $\Omega\succeq\Upsilon$, $\Omega^{-1}\pe\Upsilon^{-1}$, which gives the second bound. For the first, let $k=\rank(A)$. If $k=0$ then $P_A=0$, and if $k=n$ then $P_A=I$ and $\Omega\succeq\upsilon A^TA$. Otherwise let $O=(O_1\ O_2)$ be an orthogonal matrix whose first $k$ columns span the row space of $A$ and whose last $n-k$ columns span $\mathcal N(A)$, so that $P_A=O_1O_1^T$ and $(A^TA)^+=O_1B_A^{-1}O_1^T$ with $B_A=O_1^TA^TAO_1$ positive definite. With $\Upsilon_{ab}=O_a^T\Upsilon O_b$ for $a,b\in\{1,2\}$,
\[
O^T\Omega O=\begin{pmatrix}\upsilon B_A+\Upsilon_{11}&\Upsilon_{12}\\ \Upsilon_{21}&\Upsilon_{22}\end{pmatrix},
\]
and by the formula for the inverse of a partitioned matrix $O_1^T\Omega^{-1}O_1=(\upsilon B_A+\Upsilon_{11}-\Upsilon_{12}\Upsilon_{22}^{-1}\Upsilon_{21})^{-1}$. Since $O^T\Upsilon O$ is positive definite, its Schur complement $\Upsilon_{11}-\Upsilon_{12}\Upsilon_{22}^{-1}\Upsilon_{21}$ is positive definite, so $O_1^T\Omega^{-1}O_1\pe(\upsilon B_A)^{-1}$ and $P_A\Omega^{-1}P_A=O_1(O_1^T\Omega^{-1}O_1)O_1^T\pe\upsilon^{-1}(A^TA)^+$. If $Q\succeq\Omega$ then $Q^{-1}\pe\Omega^{-1}$, and both bounds carry over.

(b) Since $P_A+\Pi_A=I$ and $\tr(P_A\Omega^{-1}\Pi_A)=\tr(\Omega^{-1}\Pi_AP_A)=0$, we have $\tr(\Omega^{-1})=\tr(P_A\Omega^{-1}P_A)+\tr(\Pi_A\Omega^{-1}\Pi_A)$. Part (a) bounds these by $\upsilon^{-1}\tr\{(A^TA)^+\}$ and $\tr(\Pi_A\Upsilon^{-1}\Pi_A)=\sum_i\ell_i\upsilon_i^{-1}$.

(c) Let $\upsilon_{\min}=\min_i\upsilon_i$ and let $\vartheta_1,\dots,\vartheta_n$ be the eigenvalues of $A^TA$. Since $\Omega\succeq\upsilon A^TA+\upsilon_{\min}I$,
\[
\tr(A\Omega^{-1}A^T)\le\tr\{A(\upsilon A^TA+\upsilon_{\min}I)^{-1}A^T\}=\sum_{i:\vartheta_i>0}\frac{\vartheta_i}{\upsilon\vartheta_i+\upsilon_{\min}}\le\frac{\rank(A)}{\upsilon}.\qedhere
\]
\end{proof}

Since $\sum_i\ell_i=n-\rank(A)$, part (b) implies $\tr(\Omega^{-1})\le\upsilon^{-1}\tr\{(A^TA)^+\}+(n-\rank(A))\upsilon_{\min}^{-1}$, which is the bound of \citet{roma:hobe:2014}. Part (a) is what the bound on the term $V_1$ in the proof of Theorem~\ref{thm:general} uses: it bounds the variance of each coordinate of the projection onto $\mathcal N(A)$ by the weights $(\Pi_A)_{ij}^2$.

We now collect bounds for the conditional moments of $\theta$ given $(\lambda,\eta)$. Let $\tilde X=X\Sigma_\beta^{1/2}$ and let $\varsigma_{\max}$ be its largest singular value. Since $T_\lambda=\Sigma_\beta^{-1/2}(\lambda_e\tilde X^T\tilde X+I)\Sigma_\beta^{-1/2}$, we have $\lambda_eXT_\lambda^{-1}X^T=\lambda_e\tilde X(\lambda_e\tilde X^T\tilde X+I)^{-1}\tilde X^T$. A singular value decomposition $\tilde X=U\diag(\varsigma_i)V^T$ shows that the eigenvalues of this matrix are $\lambda_e\varsigma_i^2/(1+\lambda_e\varsigma_i^2)$ for the singular values $\varsigma_i$ of $\tilde X$, and zero otherwise. Hence the eigenvalues of $M_\lambda$ are $(1+\lambda_e\varsigma_i^2)^{-1}$ and $1$, and
\begin{equation}\label{eq:Mbounds}
(1+\lambda_e\varsigma_{\max}^2)^{-1}I\pe M_\lambda\pe I,\qquad XT_\lambda^{-1}X^T\pe X\Sigma_\beta X^T,
\end{equation}
the second because $T_\lambda\succeq\Sigma_\beta^{-1}$. Let $c_\lambda^{-1}=\lambda_e^{-1}+\varsigma_{\max}^2$, so that $\lambda_eZ^TM_\lambda Z\succeq c_\lambda Z^TZ$ and $Q_{\lambda,\eta}\succeq c_\lambda Z^TZ+\lambda_u\Deta$.

\begin{lemma}\label{lem:traces}
For all $\lambda\in\R_+^2$ and $\eta\in\R_+^q$,
\begin{enumerate}[label=\textup{(\alph*)},leftmargin=*]
\item $\tr(\Deta^{1/2}Q_{\lambda,\eta}^{-1}\Deta^{1/2})\le c_\lambda^{-1}\tau_Z\max_i\eta_i+(q-r)\lambda_u^{-1}$;
\item $\tr\{\Var(W\theta\mid\lambda,\eta)\}\le c_\lambda^{-1}r+\tr(X\Sigma_\beta X^T)$.
\end{enumerate}
\end{lemma}

\begin{proof}
(a) $\Deta^{1/2}Q_{\lambda,\eta}^{-1}\Deta^{1/2}=(\Deta^{-1/2}Q_{\lambda,\eta}\Deta^{-1/2})^{-1}\pe(c_\lambda\Deta^{-1/2}Z^TZ\Deta^{-1/2}+\lambda_uI)^{-1}$. Lemma~\ref{lem:trace}(b) with $A=Z\Deta^{-1/2}$, whose null space has dimension $q-r$, gives the bound $c_\lambda^{-1}\tr\{(A^TA)^+\}+(q-r)\lambda_u^{-1}$. Now $\tr\{(A^TA)^+\}=\tr\{(AA^T)^+\}=\tr\{(Z\Deta^{-1}Z^T)^+\}$. The matrices $Z\Deta^{-1}Z^T$ and $ZZ^T$ have the same range and $Z\Deta^{-1}Z^T\succeq(\max_i\eta_i)^{-1}ZZ^T$; restricting both to their common range, where they are positive definite, and inverting gives $(Z\Deta^{-1}Z^T)^+\pe\max_i\eta_i\,(ZZ^T)^+$. Since $\tr\{(ZZ^T)^+\}=\tau_Z$, (a) follows.

(b) Given $(\lambda,\eta,u)$, $\beta$ is normal with precision $T_\lambda$ and mean $T_\lambda^{-1}\{\lambda_eX^T(y-Zu)+\Sigma_\beta^{-1}\mu_\beta\}$, so $\E(X\beta+Zu\mid\lambda,\eta,u)$ equals a constant plus $M_\lambda Zu$. By the law of total variance,
\[
\Var(W\theta\mid\lambda,\eta)=XT_\lambda^{-1}X^T+M_\lambda ZQ_{\lambda,\eta}^{-1}Z^TM_\lambda .
\]
By \eqref{eq:Mbounds}, $\tr(XT_\lambda^{-1}X^T)\le\tr(X\Sigma_\beta X^T)$ and, since $0\pe M_\lambda^2\pe I$, $\tr(M_\lambda ZQ_{\lambda,\eta}^{-1}Z^TM_\lambda)\le\tr(ZQ_{\lambda,\eta}^{-1}Z^T)\le\tr\{Z(c_\lambda Z^TZ+\lambda_u\Deta)^{-1}Z^T\}\le c_\lambda^{-1}r$ by Lemma~\ref{lem:trace}(c).
\end{proof}

\begin{lemma}\label{lem:means}
There are finite constants $K_1$ and $K_2$, depending only on $X$, $Z$, $y$, $\mu_\beta$ and $\Sigma_\beta$, such that for all $\lambda\in\R_+^2$ and $\eta\in\R_+^q$,
\[
\norm{\E(\theta\mid\lambda,\eta)}\le K_1,\qquad\norm{y-W\E(\theta\mid\lambda,\eta)}\le K_2 .
\]
The same constants apply when $\lambda_u\Deta$ is replaced by any positive diagonal matrix.
\end{lemma}

\begin{proof}
Since $\theta\mid\lambda,\eta$ is normal, its mean is the maximizer of its full conditional density, that is, the minimizer of
\[
\lambda_e\norm{y-W\theta}^2+\lambda_u\sum_{i=1}^q\eta_iu_i^2+(\beta-\mu_\beta)^T\Sigma_\beta^{-1}(\beta-\mu_\beta)=\norm{D^{1/2}(b-A\theta)}^2,
\]
where
\[
A=\begin{pmatrix}X&Z\\0&I_q\\\Sigma_\beta^{-1/2}&0\end{pmatrix},\qquad b=\begin{pmatrix}y\\0\\\Sigma_\beta^{-1/2}\mu_\beta\end{pmatrix},\qquad D=\diag(\lambda_e1_N,\lambda_u\eta_1,\dots,\lambda_u\eta_q,1_p).
\]
The last $q+p$ rows of $A$ form an invertible matrix, so $A$ has full column rank, and $\E(\theta\mid\lambda,\eta)=(A^TDA)^{-1}A^TDb=A^+\{A(A^TDA)^{-1}A^TD\}b$, where $A^+=(A^TA)^{-1}A^T$ satisfies $A^+A=I$. By Lemma~\ref{lem:stewart} in Appendix~\ref{app:KH}, $\chi_A=\sup_D\norm{A(A^TDA)^{-1}A^TD}_2<\infty$, the supremum being over all positive diagonal matrices $D$. Hence $\norm{\E(\theta\mid\lambda,\eta)}\le K_1:=\norm{A^+}_2\chi_A\norm{b}$, which bounds $\norm{\E(u\mid\lambda,\eta)}$, and $\norm{y-W\E(\theta\mid\lambda,\eta)}\le K_2:=\norm{y}+\norm{W}_2K_1$. The argument uses only that the weights of the rows of $A$ are positive, which proves the last statement.
\end{proof}

\subsection{The tilted family and elementary inequalities}

\begin{lemma}\label{lem:tilted}
Assume $m_{1/2}<\infty$.
\begin{enumerate}[label=\textup{(\alph*)},leftmargin=*]
\item If $0\le s<s'$, then $H_{s'}$ is smaller than $H_s$ in the likelihood ratio order. Consequently, for every non-decreasing $g\ge0$, $s\mapsto\E_s[g(\eta)]$ is non-increasing on $[0,\infty)$, and for every non-increasing $g\ge0$ it is non-decreasing. In particular $M_c=\E_0[\eta^c]=m_{c+1/2}/m_{1/2}$ for $c>0$.
\item Let $c\in(0,1]$ with $\kappa_H(c)<\infty$. For every $\kappa'>\kappa_H(c)$ there is $C<\infty$ such that $\E_s[\eta^{-c}]\le\kappa's^c+C$ for all $s\ge0$.
\end{enumerate}
\end{lemma}

\begin{proof}
(a) The density of $H_{s'}$ with respect to $H_s$ is $w(\eta)/\E_s[w]$ with $w(\eta)=e^{-(s'-s)\eta}$, which is non-increasing in $\eta$; this is the likelihood ratio order. For a bounded non-decreasing $g$,
\[
\E_{s'}[g]-\E_s[g]=\frac{\E_s[gw]-\E_s[g]\E_s[w]}{\E_s[w]}=\frac{\mathrm{Cov}_s(g,w)}{\E_s[w]}\le0,
\]
because the covariance of a non-decreasing and a non-increasing function of the same random variable is non-positive (Chebyshev's association inequality). For a non-decreasing $g\ge0$, apply this to $\min(g,n)$ and let $n\to\infty$ by monotone convergence; the conclusion holds with infinite expectations allowed. For non-increasing $g\ge0$ the inequality is reversed. In particular the supremum defining $M_c$ is attained at $s=0$, where $\E_0[\eta^c]=m_{c+1/2}/m_{1/2}$.

(b) Choose $S>0$ such that $\E_s[\eta^{-c}]\le\kappa's^c$ for $s\ge S$. Since $\eta^{-c}$ is non-increasing, part (a) gives $\E_s[\eta^{-c}]\le\E_S[\eta^{-c}]\le\kappa'S^c=:C$ for $0\le s<S$.
\end{proof}

\begin{lemma}\label{lem:elem}
Let $c\in(0,1]$ and $\mu_c=\E|\xi|^{2c}=2^c\Gamma(c+\frac12)/\Gamma(\frac12)$ for $\xi\sim\N(0,1)$.
\begin{enumerate}[label=\textup{(\alph*)},leftmargin=*]
\item For $p>0$ and $\epsilon>0$ there is $C_{\epsilon,p}<\infty$ with $(x+y)^p\le(1+\epsilon)x^p+C_{\epsilon,p}y^p$ for all $x,y\ge0$.
\item For $w_j\ge0$, $(\sum_jw_j)^c\le\sum_jw_j^c$.
\item If $Y\sim\N(m,\sigma^2)$, then $\E|Y|^{2c}\le(1+\epsilon)\mu_c\sigma^{2c}+C_{\epsilon,2c}|m|^{2c}$ for every $\epsilon>0$.
\item If $G\sim\Ga(A,B)$ with $A>c$, then $\E[G^{-c}]=B^c\,\Gamma(A-c)/\Gamma(A)$ and $\E[G^c]\le(A/B)^c$.
\item Let $x\ge0$, $A>1$ and $f(c)=(x/2)^c\,\Gamma(A-c)/\Gamma(A)$. If $f(1)<1$, then $f(c)<1$ for every $c\in(0,1]$.
\end{enumerate}
\end{lemma}

\begin{proof}
(a) For $p\le1$, $t\mapsto t^p$ is subadditive, and one can take $C_{\epsilon,p}=1$. For $p>1$ and $t\in(0,1)$, convexity gives $(x+y)^p=\{(1-t)\frac{x}{1-t}+t\frac yt\}^p\le(1-t)^{1-p}x^p+t^{1-p}y^p$; choose $t$ with $(1-t)^{1-p}\le1+\epsilon$. (b) is subadditivity of $t\mapsto t^c$. (c) Write $Y=m+\sigma\xi$; by (a) with $p=2c$, $|Y|^{2c}\le(1+\epsilon)\sigma^{2c}|\xi|^{2c}+C_{\epsilon,2c}|m|^{2c}$. (d) The first formula is a direct integration, and the second is Jensen's inequality. (e) If $x=0$ there is nothing to prove. Otherwise $\log f(c)=c\log(x/2)+\log\Gamma(A-c)-\log\Gamma(A)$ is convex on $[0,1]$, because $\log\Gamma$ is convex, and $\log f(0)=0$; hence $\log f(c)\le c\log f(1)<0$.
\end{proof}

With $x=r$ and $A=a_e+N/2$, $f(1)=r/(2a_e+N-2)$, and $f(1)<1$ is \ref{C1}. With $x=q-r$ and $A=a_u+q/2$, $f(1)=(q-r)/(2a_u+q-2)$, and $f(1)<1$ is \ref{C2}. Moreover \ref{C1} gives $a_e+N/2>1+r/2\ge1$, and, when $r<q$, \ref{C2} gives $a_u+q/2>1+(q-r)/2>1$.

\subsection{Proof of Theorem~\ref{thm:general}}

Fix $c\in(0,1]$ as in \ref{H1}--\ref{H2}. Then $m_{1/2}\le1+m_{c+1/2}<\infty$, so Lemmas~\ref{lem:topology} and~\ref{lem:tilted} apply. Fix $(\eta',\theta')$ with $\theta'=(\beta',u')$, and write $\E'$ for conditional expectation given $(\eta',\theta')$ under one step of the marginal chain. We evaluate expectations by iterating $\E'[\,\cdot\,]=\E'\{\E[\E(\,\cdot\mid\lambda,\eta)\mid\lambda,\theta']\}$. Let $\epsilon>0$ and $\kappa'=\kappa_H(c)+\epsilon$. By \ref{H2} we may choose $\epsilon$ so small that
\[
\bar\gamma:=(1+\epsilon)^2\,\frac{\Gamma(c+\frac12)}{\Gamma(\frac12)}\,\kappa'\,S_Z(c)<1 .
\]
Consider
\[
v(\eta,\theta)=\alpha_1\sum_{i=1}^q|u_i|^{2c}+\alpha_2\norm{\Deta^{1/2}u}^{2c}+\alpha_3\norm{y-W\theta}^{2c}+\sum_{i=1}^q(\eta_i^c+\eta_i^{-c})+\norm{\beta}^{2c},
\]
with $\alpha_1,\alpha_3>0$ to be chosen, $\alpha_2>0$ to be chosen if $r<q$, and $\alpha_2=0$ if $r=q$. Write $V_1,\dots,V_5$ for the five terms without their coefficients and $V_k'$ for their values at $(\eta',\theta')$. Since $\eta^c+\eta^{-c}\ge2$, $v\ge1$. Below, $C_1,C_2,\dots$ and $L_1,L_2,\dots$ denote finite constants that do not depend on $(\eta',\theta')$.

\emph{Sublevel sets are compact.} Let $\bar v\ge1$. If $v(\eta,\theta)\le\bar v$, then $\eta_i^c+\eta_i^{-c}\le\bar v$, so each $\eta_i$ lies in the compact interval $[\bar v^{-1/c},\bar v^{1/c}]\subset\R_+$. Moreover $|u_i|^{2c}\le\bar v/\alpha_1$ for every $i$ and $\norm\beta^{2c}\le\bar v$. As $v$ is continuous, $\{v\le\bar v\}$ is closed in $\mathsf X$. It is contained in $([\bar v^{-1/c},\bar v^{1/c}]\cap\mathsf S)^q\times\bar B$ for a closed ball $\bar B\subset\R^{p+q}$, a set that is closed in $\R^{2q+p}$ because $\mathsf S$ is closed in $\R_+$. Hence $\{v\le\bar v\}$ is closed and bounded in $\R^{2q+p}$, that is, compact.

\emph{Moments of the scales and the weights.} Let $A_e=a_e+N/2$ and $G_e=\Gamma(A_e-c)/\Gamma(A_e)$, which is finite because $A_e>1\ge c$ by \ref{C1}. When $r<q$, let also $A_u=a_u+q/2$ and $G_u=\Gamma(A_u-c)/\Gamma(A_u)$, which is finite because $A_u>1$ by \ref{C2}. By Lemma~\ref{lem:elem}(d) and (b),
\begin{equation}\label{eq:lambdamoments}
\E'[\lambda_e^{-c}]\le G_e\big(b_e^c+2^{-c}V_3'\big),\qquad
\E'[\lambda_u^{-c}]\le G_u\big(b_u^c+2^{-c}V_2'\big)\ \text{ if }r<q,\qquad
\E'[\lambda_u^c]\le\zeta^c:=\Big(\frac{2a_u+q}{2b_u}\Big)^c .
\end{equation}
Also $c_\lambda^{-c}\le\lambda_e^{-c}+\varsigma_{\max}^{2c}$. Given $(\lambda,\theta')$, the $\eta_j$ are independent with laws $H_{s_j}$, $s_j=\lambda_uu_j'^2/2$, which do not involve $\lambda_e$. By Lemma~\ref{lem:tilted},
\begin{equation}\label{eq:etamoments}
\E[\eta_j^c\mid\lambda,\theta']\le M_c,\qquad
\lambda_u^{-c}\E[\eta_j^{-c}\mid\lambda,\theta']\le\kappa'2^{-c}|u_j'|^{2c}+C_0\lambda_u^{-c},
\end{equation}
where $M_c<\infty$ by \ref{H1} and $C_0$ is the constant $C$ of Lemma~\ref{lem:tilted}(b) for $\kappa'$. The cancellation of $\lambda_u$ in the second bound is the key step of the argument.

\emph{The term $V_1$.} Given $(\lambda,\eta)$, $u_i\sim\N(m_i,\sigma_i^2)$ with $m=\E(u\mid\lambda,\eta)$, $|m_i|\le K_1$ by Lemma~\ref{lem:means}, and $\sigma_i^2=e_i^TQ_{\lambda,\eta}^{-1}e_i$, where $e_i$ is the $i$th coordinate vector. Write $\Pi=\Pi_Z$ and $P_Z=I-\Pi$. By the triangle inequality, $\sigma_i=\norm{Q_{\lambda,\eta}^{-1/2}e_i}\le\norm{Q_{\lambda,\eta}^{-1/2}\Pi e_i}+\norm{Q_{\lambda,\eta}^{-1/2}P_Ze_i}$. Lemma~\ref{lem:trace}(a), with $A=Z$, $\upsilon=c_\lambda$, $\Upsilon=\lambda_u\Deta$ and $Q=Q_{\lambda,\eta}$, gives
\[
e_i^T\Pi Q_{\lambda,\eta}^{-1}\Pi e_i\le\sum_j\Pi_{ij}^2(\lambda_u\eta_j)^{-1},\qquad
e_i^TP_ZQ_{\lambda,\eta}^{-1}P_Ze_i\le c_\lambda^{-1}\{(Z^TZ)^+\}_{ii}.
\]
By Lemma~\ref{lem:elem}(a) with $p=2c$ and Lemma~\ref{lem:elem}(b),
\[
\sigma_i^{2c}\le(1+\epsilon)\sum_j|\Pi_{ij}|^{2c}\lambda_u^{-c}\eta_j^{-c}+C_1c_\lambda^{-c},
\]
Lemma~\ref{lem:elem}(c) then gives
\[
\E(V_1\mid\lambda,\eta)\le(1+\epsilon)^2\mu_c\sum_i\sum_j|\Pi_{ij}|^{2c}\lambda_u^{-c}\eta_j^{-c}+C_2c_\lambda^{-c}+C_3 ,
\]
with $C_2=(1+\epsilon)\mu_cqC_1$ and $C_3=qC_{\epsilon,2c}K_1^{2c}$. Using \eqref{eq:etamoments}, the identity $\mu_c2^{-c}=\Gamma(c+\frac12)/\Gamma(\frac12)$ and $\sum_i|\Pi_{ij}|^{2c}\le S_Z(c)$, we obtain
\[
\E(V_1\mid\lambda,\theta')\le\bar\gamma\sum_j|u_j'|^{2c}+C_4\lambda_u^{-c}\mathbf 1\{r<q\}+C_2c_\lambda^{-c}+C_3,
\]
where $C_4=(1+\epsilon)^2\mu_cqS_Z(c)C_0$ and the term in $\lambda_u^{-c}$ is present only if $\Pi\ne0$, that is, if $r<q$. By \eqref{eq:lambdamoments},
\[
\E'V_1\le\bar\gamma V_1'+A_{12}V_2'+A_{13}V_3'+L_1,
\]
with $A_{12}=C_4G_u2^{-c}$ if $r<q$, $A_{12}=0$ if $r=q$, and $A_{13}=C_2G_e2^{-c}$.

\emph{The term $V_2$ (only if $r<q$).} By Lemmas~\ref{lem:traces}(a) and~\ref{lem:means}, and since $\max_i\eta_i\le\sum_i\eta_i$,
\[
\E(\norm{\Deta^{1/2}u}^2\mid\lambda,\eta)\le\big(c_\lambda^{-1}\tau_Z+K_1^2\big)\sum_i\eta_i+(q-r)\lambda_u^{-1}.
\]
By Jensen's inequality and Lemma~\ref{lem:elem}(b), $\E(V_2\mid\lambda,\eta)\le(c_\lambda^{-c}\tau_Z^c+K_1^{2c})\sum_i\eta_i^c+(q-r)^c\lambda_u^{-c}$. Taking expectations with \eqref{eq:etamoments} and \eqref{eq:lambdamoments},
\[
\E'V_2\le A_{22}V_2'+A_{23}V_3'+L_2,
\]
with $A_{22}=(q-r)^c2^{-c}G_u$ and $A_{23}=qM_c\tau_Z^c2^{-c}G_e$. By Lemma~\ref{lem:elem}(e) with $x=q-r$ and $A=A_u$, $A_{22}<1$ under \ref{C2}.

\emph{The term $V_3$.} By Lemmas~\ref{lem:traces}(b) and~\ref{lem:means}, $\E(\norm{y-W\theta}^2\mid\lambda,\eta)\le c_\lambda^{-1}r+\tr(X\Sigma_\beta X^T)+K_2^2$. By Jensen's inequality and Lemma~\ref{lem:elem}(b),
\[
\E'V_3\le A_{33}V_3'+L_3,\qquad A_{33}=r^c2^{-c}G_e,
\]
and $A_{33}<1$ under \ref{C1} by Lemma~\ref{lem:elem}(e) with $x=r$ and $A=A_e$.

\emph{The term $V_4$.} By \eqref{eq:etamoments} and \eqref{eq:lambdamoments},
\[
\E'V_4\le qM_c+\sum_j\big(\kappa'2^{-c}\E'[\lambda_u^c]\,|u_j'|^{2c}+C_0\big)\le\kappa'2^{-c}\zeta^cV_1'+L_4 .
\]

\emph{The term $V_5$.} Given $(\lambda,\eta)$, $\beta$ is normal with $\norm{\E(\beta\mid\lambda,\eta)}\le\norm{\E(\theta\mid\lambda,\eta)}\le K_1$ by Lemma~\ref{lem:means}, and $\Var(\beta\mid\lambda,\eta)\pe\Sigma_\beta$: indeed $P_{\lambda,\eta}=\lambda_eW^TW+\diag(\Sigma_\beta^{-1},\lambda_u\Deta)\succeq\diag(\Sigma_\beta^{-1},\lambda_u\Deta)$, so $P_{\lambda,\eta}^{-1}\pe\diag(\Sigma_\beta,(\lambda_u\Deta)^{-1})$, and $\Var(\beta\mid\lambda,\eta)$ is the leading $p\times p$ block of $P_{\lambda,\eta}^{-1}$. By Jensen's inequality, $\E(V_5\mid\lambda,\eta)\le\{\E(\norm\beta^2\mid\lambda,\eta)\}^c\le\{\tr(\Sigma_\beta)+K_1^2\}^c=:L_5$, so $\E'V_5\le L_5$. This is the only place where $\beta$ enters the drift function; no condition on the rank of $X$ is needed.

\emph{Conclusion.} Collecting terms,
\[
\E'v\le\Big(\bar\gamma+\frac{\kappa'2^{-c}\zeta^c}{\alpha_1}\Big)\alpha_1V_1'+\Big(A_{22}+\frac{\alpha_1A_{12}}{\alpha_2}\Big)\alpha_2V_2'+\Big(A_{33}+\frac{\alpha_1A_{13}+\alpha_2A_{23}}{\alpha_3}\Big)\alpha_3V_3'+L,
\]
where the second term is absent if $r=q$. Choose first $\alpha_1>\kappa'2^{-c}\zeta^c/(1-\bar\gamma)$, then (if $r<q$) $\alpha_2>\alpha_1A_{12}/(1-A_{22})$, and finally $\alpha_3>(\alpha_1A_{13}+\alpha_2A_{23})/(1-A_{33})$. Every coefficient is then smaller than one, and since $V_4',V_5'\ge0$ and $\E'V_5\le L_5$, \eqref{eq:drift} holds with some $\rho<1$. \qed

\subsection{Proof of Theorem~\ref{thm:anydesign}}\label{sec:proof-anydesign}

\emph{Step 1: coordinate-specific mixing distributions.} Consider the model in which $\eta_i\sim H_i$ independently, with possibly different laws $H_i$ satisfying $m_{1/2}(H_i)<\infty$, and let $\mathsf S_i$ be the support of $H_i$. Lemma~\ref{lem:topology} holds with the same proof, with $\varphi=\bigotimes_iH_i\otimes\mathrm{Leb}_{p+q}$ and state space $\prod_i\mathsf S_i\times\R^{p+q}$. Lemmas~\ref{lem:trace}, \ref{lem:traces} and~\ref{lem:means} do not involve the mixing distributions. In the proof of Theorem~\ref{thm:general}, replace $M_c$ by $\max_im_{c+1/2}(H_i)/m_{1/2}(H_i)$, $\kappa'$ by $\kappa_i'=\kappa_{H_i}(c)+\epsilon$, and $C_0$ by the maximum of the constants of Lemma~\ref{lem:tilted}(b) over $i$. The bound for $V_1$ then holds with $\bar\gamma=(1+\epsilon)^2\frac{\Gamma(c+\frac12)}{\Gamma(\frac12)}\max_j\big\{\kappa_j'\sum_i|(\Pi_Z)_{ij}|^{2c}\big\}$. The other bounds are unchanged, with $\kappa'$ replaced by $\max_i\kappa_i'$ in the term $V_4$. Hence the chain is geometrically ergodic under \ref{C1}--\ref{C2} if there is $c\in(0,1]$ with $m_{c+1/2}(H_i)<\infty$ and $\kappa_{H_i}(c)<\infty$ for every $i$ and
\begin{equation}\label{eq:H2coord}
R_{H_j}(c)\sum_i|(\Pi_Z)_{ij}|^{2c}<1\qquad\text{for every }j.
\end{equation}

\emph{Step 2: rescaling.} Let $C=\diag(\omega_1,\dots,\omega_q)$ with $\omega_i>0$, let $\tilde Z=ZC^{-1}$, and let $\tilde H_i$ be the law of $\eta/\omega_i^2$ when $\eta\sim H$. Consider the model of Step~1 with $Z$ replaced by $\tilde Z$, $H_i$ by $\tilde H_i$, and the same priors, and the bijection $\mathcal T(\lambda,\eta,\theta)=(\lambda,C^{-2}\eta,(\beta,Cu))$. If $u_i\mid\lambda_u,\eta_i\sim\N(0,(\lambda_u\eta_i)^{-1})$, then $\omega_iu_i\mid\lambda_u,\eta_i\sim\N(0,(\lambda_u\eta_i/\omega_i^2)^{-1})$, and $\tilde ZCu=Zu$. Hence the joint law of $(\mathcal T(\lambda,\eta,\theta),Y)$ under the original model is the joint law of the parameters and $Y$ under the rescaled model. The posterior of the rescaled model is therefore the image of the original posterior under $\mathcal T$. Since $\mathcal T$ acts separately on the three blocks, each full conditional of the rescaled model is the image of the corresponding full conditional of the original model. This can also be checked directly on the explicit formulas of Section~\ref{sec:model}, so that it holds for every state and not only almost everywhere: $\norm{y-W\theta}$ and $\sum_i\eta_iu_i^2$ are unchanged, the tilting parameters become $\tilde s_i=\omega_i^2s_i$, the law $\tilde H_i$ tilted at $\tilde s_i$ is the image of $H_{s_i}$ under $\eta\mapsto\eta/\omega_i^2$, and the normal full conditional of $\theta$ is mapped linearly. Hence the Gibbs chain of the rescaled model started at $\mathcal T(x)$ has the law of $\mathcal T(\Phi)$ started at $x$. Total variation distance is invariant under bimeasurable bijections, so geometric ergodicity of one chain is equivalent to that of the other. The ranks of $\tilde Z$ and $Z$ coincide, so \ref{C1}--\ref{C2} are unchanged, and $m_t(\tilde H_i)=\omega_i^{-2t}m_t(H)$ for every $t>0$. The law $\tilde H_i$ tilted at $s$ is the law of $\eta/\omega_i^2$ with $\eta\sim H_{s/\omega_i^2}$, so $\E_{(\tilde H_i)_s}[x^{-c}]=\omega_i^{2c}\E_{s/\omega_i^2}[\eta^{-c}]$ and $\kappa_{\tilde H_i}(c)=\kappa_H(c)$. Hence the rescaled model satisfies the conditions of Step~1 if \ref{H1} holds and $R_H(c)\sum_i|(\tilde\Pi)_{ij}|^{2c}<1$ for every $j$, where $\tilde\Pi$ is the orthogonal projection onto $\mathcal N(\tilde Z)=C\mathcal N(Z)$.

\emph{Step 3: choice of the rescaling.} If $r=q$ there is nothing to prove, so let $m=q-r>0$. Choose $r$ linearly independent columns of $Z$, let $J$ be the set of the remaining $m$ coordinates, and order the coordinates so that $J=\{1,\dots,m\}$; a permutation of the coordinates of $u$, with the corresponding permutation of the columns of $Z$ and of the weights, is a relabelling that maps the Gibbs chain to the Gibbs chain of the permuted model and leaves all the conditions unchanged. The columns outside $J$ form a basis of the column space of $Z$, so for every $v_J\in\R^m$ there is exactly one $v\in\mathcal N(Z)$ with these first $m$ coordinates. Hence $\mathcal N(Z)$ is the column space of a matrix $\begin{pmatrix}I_m\\A\end{pmatrix}$. For $M\ge1$ let $C_M=\diag(M1_m,1_{q-m})$. Then $C_M\mathcal N(Z)$ is the column space of $B_M=\begin{pmatrix}I_m\\A/M\end{pmatrix}$, whose orthogonal projection $\tilde\Pi_M=B_M(B_M^TB_M)^{-1}B_M^T$ converges entrywise to $\diag(1_m,0_{q-m})$ as $M\to\infty$, because $B_M^TB_M=I_m+A^TA/M^2\to I_m$. For fixed $c\in(0,1]$, $\sum_i|(\tilde\Pi_M)_{ij}|^{2c}$ therefore converges to $1$ for $j\in J$ and to $0$ for $j\notin J$. Since $R_H(c)<1$, the condition of Step~2 holds for $M$ large enough, and the chain is geometrically ergodic. \qed

\subsection{Proof of Proposition~\ref{prop:groups}}

Let $\Lambda_\eta=\diag(\lambda_{u,k(1)}\eta_1,\dots,\lambda_{u,k(q)}\eta_q)$. Lemma~\ref{lem:topology} holds with the same proof, taking $q_k\delta\bar u^2<b_{u,k}$ for each group $k$ in the domination step, where $q_k$ is the number of random effects in group $k$. Lemmas~\ref{lem:trace}, \ref{lem:traces}(b) and~\ref{lem:means} hold with $\lambda_u\Deta$ replaced by $\Lambda_\eta$, since their proofs only use that this matrix is positive and diagonal. Lemma~\ref{lem:tilted} concerns $H$ only. Since $r=q$, $\Pi_Z=0$, and the proof of Theorem~\ref{thm:general} applies with $\alpha_2=0$. The only change is in the term $V_4$, where $s_j=\lambda_{u,k(j)}u_j'^2/2$ and, by Jensen's inequality, $\E'[\lambda_{u,k}^c]\le\zeta_k^c$ with $\zeta_k=(2a_{u,k}+q_k)/(2b_{u,k})$. Replacing $\zeta$ by $\max_k\zeta_k$ completes the argument. \qed

\subsection{Proof of Proposition~\ref{prop:families}}

(a) Under every $H_s$, $\eta\ge\delta$ almost surely, so $\E_s[\eta^{-c}]\le\delta^{-c}$ and $\kappa_H(c)=0$.

(b) The condition on $(\varrho,\psi)$ makes $H$ a proper law with $m_{1/2}<\infty$, and $H_s=\GIG(\varrho+\frac12,\chi,\psi+2s)$, with $\psi+2s>0$ for $s>0$. For $\varrho'\in\R$, $\chi>0$ and $\psi'>0$, the mean of $\eta^{-1}$ under $\GIG(\varrho',\chi,\psi')$ is $(\psi'/\chi)^{1/2}K_{\varrho'-1}(\omega)/K_{\varrho'}(\omega)$ with $\omega=(\chi\psi')^{1/2}$, where $K_\nu$ is the modified Bessel function of the second kind \citep{jorg:1982}. Since $K_\nu(\omega)\sim(\pi/2\omega)^{1/2}e^{-\omega}$ as $\omega\to\infty$ for every fixed $\nu$ \citep[\S10.40(i)]{DLMF}, $K_{\varrho'-1}(\omega)/K_{\varrho'}(\omega)\to1$ and $\E_s[\eta^{-1}]=O(s^{1/2})$ as $s\to\infty$. By Jensen's inequality $\E_s[\eta^{-c}]=O(s^{c/2})$, so $\kappa_H(c)=0$.

(c) Let $-a<b\le\frac12$. The substitution $\eta=1/t$ turns $\int_0^x\eta^{a+b-1}L(\eta)\,d\eta$ into $\int_{1/x}^\infty t^{-(a+b)-1}L(1/t)\,dt$, where $t\mapsto L(1/t)$ is slowly varying at infinity; by Karamata's theorem for tail integrals \citep[Proposition~1.5.10]{bing:gold:teug:1987}, the measure $G_b(d\eta)=\eta^bh(\eta)\,d\eta$ on $(0,\epsilon)$ satisfies $G_b((0,x])\sim x^{a+b}L(x)/(a+b)$ as $x\downarrow0$. By Karamata's Abelian theorem for Laplace--Stieltjes transforms, in its version for $s\to\infty$ \citep[Section~XIII.5]{Feller1971},
\[
\int_0^\epsilon e^{-s\eta}\eta^bh(\eta)\,d\eta\sim\Gamma(a+b)\,s^{-(a+b)}L(1/s)\qquad(s\to\infty).
\]
For $b\le1/2$ the contribution of $[\epsilon,\infty)$ is at most $e^{-s\epsilon}\epsilon^{b-1/2}m_{1/2}$, which is $o\big(s^{-(a+b)}L(1/s)\big)$ because $L(1/s)\ge s^{-1}$ for large $s$ by Potter's bounds \citep[Theorem~1.5.6]{bing:gold:teug:1987}. Taking $b=\frac12-c$ in the numerator and $b=\frac12$ in the denominator, both of which lie in $(-a,\frac12]$,
\[
\E_s[\eta^{-c}]\sim\frac{\Gamma(a+\frac12-c)}{\Gamma(a+\frac12)}\,s^c\qquad(s\to\infty),
\]
which is the claim. \qed

\medskip
\noindent\emph{Proof of Corollary~\ref{cor:families}.} Throughout, \ref{C1}--\ref{C2} hold.

(i) $\Ga(d/2,d/2)$ has all positive moments and index $a=d/2$, so \ref{H1} holds for every $c\in(0,1]$ with $c<(d+1)/2$, and Proposition~\ref{prop:families}(c) gives $R_H(c)$ as displayed in Section~\ref{sec:results}. Since $R_H(c)\to1$ as $c\downarrow0$ and $\frac{d}{dc}\log R_H(c)$ at $c=0$ equals $\psi(\frac12)-\psi(\frac{d+1}2)<0$, where $\psi$ denotes here the digamma function, $R_H(c)<1$ for small $c>0$, and Theorem~\ref{thm:anydesign} applies for every $Z$. With $c=1$ (allowed when $d>1$), $R_H(1)=1/(d-1)$ and $S_Z(1)=\ell^*$, so \ref{H2} reads $\ell^*<d-1$.

(ii) The proof is identical, with $a=\nu$; $\mathrm{Beta}(\nu,1)$ has all positive moments because its support is bounded.

(iii) and (iv) By Proposition~\ref{prop:families}(a) and (b), $\kappa_H(c)=0$ for every $c$, so \ref{H2} holds; all moments of $H$ are finite, so \ref{H1} holds with $c=1$. The identification of the normal inverse Gaussian and hyperbolic laws follows because the law of $1/\eta$ when $\eta\sim\GIG(\varrho,\psi,\chi)$ is $\GIG(-\varrho,\chi,\psi)$.

(v) $\mathrm{IG}(\alpha,\beta)=\GIG(-\alpha,2\beta,0)$ has $m_r<\infty$ if and only if $r<\alpha$. If $\alpha>\frac12$, choose $c\in(0,1]$ with $c<\alpha-\frac12$; then \ref{H1} holds, and Proposition~\ref{prop:families}(b) gives $\kappa_H(c)=0$ and hence \ref{H2}. \qed

%% file: sec5_practice.tex
\section{Practical implications}\label{sec:practice}

The practical value of geometric ergodicity lies in the output analysis that it justifies. Let $g$ be a vector of functions of the state, let $\bar g_n=n^{-1}\sum_{k=0}^{n-1}g(\Phi^{(k)})$, and let $\E_\pi g$ denote the posterior mean. For a Harris ergodic and geometrically ergodic chain, a CLT
\[
\sqrt n\,(\bar g_n-\E_\pi g)\ \xrightarrow{d}\ \N(0,\Sigma_g)
\]
holds for every initial distribution as soon as $\E_\pi\norm{g}^{2+\delta}<\infty$ for some $\delta>0$ \citep{chan:geye:1994,Jones2004}. If $\E_\pi\norm{g}^{2+\delta+\delta'}<\infty$ for some $\delta,\delta'>0$, a multivariate strong invariance principle holds with rate $n^{\gamma_0}\log n$, $\gamma_0=\max\{1/(2+\delta),1/4\}$ \citep[Theorem~2]{bane:vats:2024}. Consequently the multivariate batch-means estimator of $\Sigma_g$ with batch size $b_n=\lfloor n^{t}\rfloor$ is strongly consistent whenever $t>\max\{2/(2+\delta),1/2\}$ \citep[Theorem~3 of][and Section~3.1 of \citealp{bane:vats:2024}]{vats:fleg:jone:2019}; see also \citet{jone:hara:caff:neat:2006} and \citet{fleg:jone:2010} for the univariate case. This justifies Monte Carlo standard errors, confidence regions for $\E_\pi g$, the multivariate effective sample size, and stopping rules based on them. The next result supplies the moment conditions for the quantities usually monitored in a mixed model.

\begin{proposition}\label{prop:moments}
For every mixing distribution $H$, under the posterior distribution,
\[
\E\norm{\beta}^k<\infty,\qquad\E|\log\lambda_e|^k<\infty,\qquad\E|\log\lambda_u|^k<\infty\qquad\text{for every }k>0 .
\]
\end{proposition}

\begin{proof}
For all $(\beta,u,\lambda_e)$, $p(y\mid\beta,u,\lambda_e)\le(\lambda_e/2\pi)^{N/2}$. Hence, for every non-negative function $F$ of the parameters,
\[
\E_\pi[F]\le\frac{(2\pi)^{-N/2}}{m(y)}\,\E_{\mathrm{prior}}\big[\lambda_e^{N/2}F\big],
\]
where $m(y)=\E_{\mathrm{prior}}[p(y\mid\beta,u,\lambda_e)]$ is the marginal likelihood, which satisfies $0<m(y)\le(2\pi)^{-N/2}\E[\lambda_e^{N/2}]<\infty$ because the likelihood is positive. Under the prior, $\beta$, $\lambda_e$ and $\lambda_u$ are independent, $\beta$ is normal and $\lambda_e,\lambda_u$ are gamma with positive shapes and rates. For $F=\norm\beta^k$ and $F=|\log\lambda_u|^k$ the right-hand side therefore factorizes, and it is finite because $\E[\lambda_e^{N/2}]=\Gamma(a_e+N/2)/\{\Gamma(a_e)b_e^{N/2}\}$, $\E\norm\beta^k$ and $\E|\log\lambda_u|^k$ are finite. For $F=|\log\lambda_e|^k$ it is a multiple of $\int_0^\infty\lambda^{a_e+N/2-1}|\log\lambda|^ke^{-b_e\lambda}\,d\lambda<\infty$. For model \eqref{eq:groupscale} the same argument applies to each $\lambda_{u,k}$.
\end{proof}

Under the conditions of Theorem~\ref{thm:general} or Theorem~\ref{thm:anydesign} (or Proposition~\ref{prop:groups}), Proposition~\ref{prop:moments} and the CLT stated above therefore apply to
\begin{equation}\label{eq:monitor}
g=(\beta^T,\log\sigma_e,\log\sigma_u)^T,\qquad\sigma_e=\lambda_e^{-1/2},\quad\sigma_u=\lambda_u^{-1/2},
\end{equation}
and, since $g$ has posterior moments of all orders, the multivariate batch-means estimator $\hat\Sigma_n$ with batch size $\lfloor n^{t}\rfloor$ is strongly consistent for every $t\in(\frac12,1)$. We use this $g$ and $t=0.6$ in the rest of the paper. The CLT also holds for bounded functions of the state, such as indicators used to estimate posterior probabilities, and for $\beta\beta^T$, so that Monte Carlo standard errors for the posterior standard deviations of the components of $\beta$ follow from the delta method. For the scales on their natural scale, the bound in the proof gives $\E_\pi[\sigma_e^k]<\infty$ for $k<2a_e+N$ and $\E_\pi[\sigma_u^k]<\infty$ for $k<2a_u$; the second condition, which is generally conservative, yields a CLT for $\sigma_u$ only when $a_u>1$, and this is why $g$ contains the logarithms of the scales. Posterior quantiles \citep[see][]{doss:fleg:jone:neat:2014} and functions of the latent weights are not covered by Proposition~\ref{prop:moments}.

The multivariate effective sample size is
\[
\mathrm{mESS}=n\,\Big\{\frac{\det(\hat\Lambda_n)}{\det(\hat\Sigma_n)}\Big\}^{1/(p+2)},
\]
where $\hat\Lambda_n$ is the sample covariance matrix of the draws. \citet{vats:fleg:jone:2019} show that, when $\hat\Sigma_n$ is strongly consistent, stopping the simulation when $\mathrm{mESS}\ge\mathrm{minESS}(p+2,\alpha,\epsilon)$ produces, asymptotically as $\epsilon\to0$, a $100(1-\alpha)\%$ confidence region for $\E_\pi g$ whose volume, raised to the power $1/(p+2)$, is a fraction $\epsilon$ of $\det(\Lambda)^{1/\{2(p+2)\}}$, where $\Lambda$ is the posterior covariance matrix of $g$. Geometric ergodicity and Proposition~\ref{prop:moments} supply the hypotheses under which this rule is valid for the sampler of Algorithm~\ref{alg:gibbs}. The region is $\{\mu:n(\bar g_n-\mu)^T\hat\Sigma_n^{-1}(\bar g_n-\mu)\le k\}$ with $k=\chi^2_{p+2,1-\alpha}$, and minESS is derived with this quantile. With $a$ batches, the Hotelling-type quantile $k_H(a)=(p+2)(a-1)/(a-p-2)\,F_{p+2,a-p-2,1-\alpha}$, which treats $\hat\Sigma_n$ as a Wishart matrix with $a-1$ degrees of freedom, accounts approximately for the variability of $\hat\Sigma_n$; the approximation is not exact, because batch means are neither exactly independent nor exactly normal. A rule that controls the volume of this corrected region stops when $\mathrm{mESS}\ge\mathrm{minESS}(p+2,\alpha,\epsilon)\,k_H(a)/\chi^2_{p+2,1-\alpha}$, with $a$ updated at each check; the factor tends to one as $a$ grows. In the application of Section~\ref{sec:app}, $p+2=7$, and with $\alpha=\epsilon=0.05$ one obtains $\mathrm{minESS}=8769$; for the models with two group-specific scales, $g$ has eight components and $\mathrm{minESS}=8804$.

%% file: sec6_sim.tex
\section{Simulation studies}\label{sec:sim}

We report three experiments. The first examines what is at stake when the random-effects distribution is misspecified, the second checks the Monte Carlo output analysis of Section~\ref{sec:practice} at finite run lengths, and the third illustrates Theorem~\ref{thm:anydesign} on a rank-deficient design. The first two use the design of the $\beta$-HCG study analysed in Section~\ref{sec:app}: $N=375$ measurements on $q=173$ women, a random intercept for each woman ($Z$ is the subject indicator matrix, of full column rank), and five fixed effects, namely a quadratic time trend, a group effect $\beta_4$ and a group-by-time interaction $\beta_5$. All three use the priors of that section.

\subsection{Misspecification of the random-effects distribution}

In the first experiment we set $\beta=(0.45,1.13,-0.07,0.30,-0.175)^T$ and $\sigma_e=0.37$, close to the estimates obtained in Section~\ref{sec:app}, and generated the random intercepts from three distributions: (i) normal with standard deviation $0.30$; (ii) Student-$t$ with 3 degrees of freedom and scale $0.22$; (iii) a contaminated normal in which each subject has, with probability $0.1$, a standard deviation five times larger than the baseline value $0.25$. For each scenario we generated 100 data sets and fitted four models: normal, Student-$t_4$, slash with $\nu=1$ and contaminated normal with $(\nu,\gamma)=(0.1,0.1)$. The tail parameters of the fitted models were held fixed, so that none of them coincides with the data-generating distribution in scenarios (ii) and (iii). Each fit used $6000$ iterations of Algorithm~\ref{alg:gibbs}, of which the first $1000$ were discarded as burn-in. By Theorem~\ref{thm:general} the Gibbs chains of all four fitted models are geometrically ergodic.

Table~\ref{tab:sim} reports, for the group effect $\beta_4$, the bias, root mean squared error (RMSE) and empirical coverage and mean width of the 95\% equal-tailed credible intervals; the RMSE of the posterior means of the random effects; and the proportion of data sets in which each model attained the smallest WAIC \citep{Watanabe2010}. WAIC was computed from the subject-level marginal likelihoods described in Appendix~\ref{app:code}. With 100 replications, the simulation standard error of a coverage near $0.95$ is about $0.02$. Those of the bias and of the RMSE of $\beta_4$ are about $1.5$ and $1.0$ to $1.6$ (on the scale of Table~\ref{tab:sim}), respectively. Comparisons between fitted models are made on the same data sets, and we report bootstrap standard errors of the paired ratios. The three scenarios have different random-effect variances (standard deviations $0.30$, $0.38$ and $0.46$), so comparisons are meaningful within each scenario rather than across scenarios. Within each fit, the Monte Carlo error of the posterior mean of $\beta_4$ is about $0.003$, which is negligible compared with its posterior standard deviation of about $0.15$.

\begin{table}[t]
\centering\small
\caption{Simulation results (100 data sets per scenario). Bias, RMSE and interval width for $\beta_4$, and RMSE of the predicted random effects ($u$-RMSE), are multiplied by 100. ``Cov.'' is the coverage (\%) of 95\% credible intervals for $\beta_4$; ``WAIC'' is the percentage of data sets in which the model has the smallest WAIC.}
\label{tab:sim}
\begin{tabular}{llrrrrrr}
\toprule
True distribution & Fitted model & Bias & RMSE & Cov. & Width & $u$-RMSE & WAIC\\
\midrule
Normal & Normal & 2.3 & 14.9 & 94 & 56.9 & 20.4 & 83 \\
 & $t_4$ & 2.1 & 14.8 & 93 & 57.1 & 20.6 & 14 \\
 & Slash(1) & 2.0 & 14.8 & 94 & 57.5 & 20.7 & 0 \\
 & CN(0.1,0.1) & 2.1 & 14.8 & 93 & 56.9 & 20.7 & 3 \\
\addlinespace
Student-$t_3$ & Normal & 0.9 & 16.0 & 92 & 60.3 & 22.0 & 9 \\
 & $t_4$ & 0.6 & 14.7 & 95 & 57.8 & 20.5 & 26 \\
 & Slash(1) & 0.5 & 14.5 & 95 & 57.6 & 20.5 & 28 \\
 & CN(0.1,0.1) & 0.5 & 14.6 & 95 & 57.7 & 20.5 & 37 \\
\addlinespace
Contaminated normal & Normal & -1.6 & 16.8 & 95 & 65.1 & 23.9 & 0 \\
 & $t_4$ & -2.5 & 14.7 & 94 & 59.6 & 21.3 & 1 \\
 & Slash(1) & -2.5 & 14.7 & 94 & 58.9 & 21.0 & 62 \\
 & CN(0.1,0.1) & -2.2 & 14.8 & 94 & 59.3 & 21.1 & 37 \\
\bottomrule
\end{tabular}
\end{table}

Three conclusions emerge. First, when the random effects are normal, the heavy-tailed models perform like the normal one: the RMSE, coverage and interval width for $\beta_4$ and the prediction error of the random effects differ little across the four fitted models, and WAIC selects the normal model in 83\% of the data sets. Second, when the random effects have heavier tails than the normal, the normal model remains approximately unbiased and its intervals keep their nominal coverage, in agreement with the robustness results of \citet{Verbeke1997} and \citet{McCulloch2011SS}, but it is less efficient. Its RMSE for $\beta_4$ is about 9\% larger than that of the heavy-tailed fits under the $t_3$ scenario (paired ratios $1.09$ to $1.11$, bootstrap standard error $0.03$) and about 14\% larger under the contaminated normal scenario (ratios $1.13$ to $1.14$, standard error $0.05$). Its credible intervals are 4--5\% and 9--11\% wider (standard errors about $0.005$), and the random effects are predicted with 7--14\% larger error. Third, the three heavy-tailed models behave very similarly to each other, so the choice among them matters less than the decision to allow heavy tails at all. WAIC never selected the normal model in the contaminated scenario and selected it in only 9\% of the data sets under the $t_3$ scenario.

In the scenarios studied, the heavy-tailed models behaved like the normal model when the random effects were normal and improved efficiency under the heavy-tailed alternatives. The next experiment turns to the precision of the Monte Carlo estimates on which such comparisons rest.

\subsection{Coverage of Monte Carlo confidence intervals}\label{sec:coverage}

For the Student-$t$ model with $d=1$, whose random effects have no mean, and for the contaminated normal model with $(\nu,\gamma)=(0.2,0.03)$, both fitted to the $\beta$-HCG data, we computed reference values of $\E_\pi g$, with $g$ as in \eqref{eq:monitor}, from an independent run of $2\times10^6$ iterations after a burn-in of $10{,}000$; their largest Monte Carlo standard error is $3\times10^{-4}$. We then ran 200 chains per model and design from dispersed starting values, discarding the first $1000$ iterations, and computed for each chain 95\% confidence intervals for the components of $\E_\pi g$ and a 95\% confidence region for the vector, with batch size $\lfloor n^{0.6}\rfloor$ and the two calibrations of Section~\ref{sec:practice}, $\chi^2_{7,0.95}$ and $k_H(a)$. The designs are two fixed run lengths, $n=20{,}000$ ($a=52$ batches) and $n=100{,}000$ ($a=100$), and two sequential rules, both checking every $1000$ iterations from $5000$ on: the rule of \citet{vats:fleg:jone:2019}, which stops when $\mathrm{mESS}\ge\mathrm{minESS}(7,0.05,0.05)=8769$ and controls the volume of the $\chi^2$ region, and the rule that stops when $\mathrm{mESS}\ge8769\,k_H(a)/\chi^2_{7,0.95}$ and controls the volume of the corrected region; for $a=52$ the second threshold is about $10{,}985$. Enlarging the region after the first rule has stopped multiplies its volume, raised to the power $1/7$, by $\{k_H(a)/\chi^2_{7,0.95}\}^{1/2}$, which is $1.12$ for $a=52$, so that rule no longer ensures that the enlarged region meets the precision target. The two sequential rules are applied to the same chains and differ only in the stopping time.

\begin{table}[t]
\centering\small\setlength{\tabcolsep}{4pt}
\caption{Coverage of 95\% Monte Carlo confidence intervals and regions for $\E_\pi g$, $g=(\beta^T,\log\sigma_e,\log\sigma_u)^T$, $\beta$-HCG data, 200 chains per row. ``Marginal'' is the range over the seven components; the joint region uses the $\chi^2_7$ or the Hotelling-type quantile $k_H(a)$. ``Sequential, $\chi^2$'' stops when $\mathrm{mESS}\ge8769$, and ``sequential, $k_H$'' when $\mathrm{mESS}\ge8769\,k_H(a)/\chi^2_{7,0.95}$; for these rows, $n$ is the median stopping time, with the 5\% and 95\% quantiles in parentheses. The simulation standard error of a coverage near $0.95$ is about $0.015$, and near $0.85$ about $0.025$.}
\label{tab:coverage}
\begin{tabular}{llrrrr}
\toprule
Model & Run length $n$ & $a$ & Marginal & Joint, $\chi^2$ & Joint, $k_H(a)$\\
\midrule
Student-$t$, $d=1$ & $20{,}000$ & 52 & 0.925--0.955 & 0.885 & 0.940\\
 & $100{,}000$ & 100 & 0.905--0.960 & 0.900 & 0.940\\
 & sequential, $\chi^2$: $19{,}000$ $(16{,}000$--$21{,}000)$ & 46--54 & 0.900--0.960 & 0.810 & 0.915\\
 & sequential, $k_H$: $23{,}000$ $(21{,}000$--$26{,}000)$ & 53--58 & 0.915--0.950 & 0.850 & 0.930\\
\addlinespace
CN$(0.2,0.03)$ & $20{,}000$ & 52 & 0.935--0.965 & 0.870 & 0.965\\
 & $100{,}000$ & 100 & 0.915--0.960 & 0.905 & 0.925\\
 & sequential, $\chi^2$: $19{,}000$ $(17{,}000$--$21{,}000)$ & 47--55 & 0.915--0.970 & 0.850 & 0.920\\
 & sequential, $k_H$: $23{,}000$ $(21{,}000$--$26{,}000)$ & 53--58 & 0.895--0.970 & 0.880 & 0.925\\
\bottomrule
\end{tabular}
\end{table}

Table~\ref{tab:coverage} shows the results. At fixed run lengths the componentwise intervals cover between $90.5\%$ and $96.5\%$ of the time, and the joint region calibrated with $\chi^2_{7,0.95}$ undercovers, with coverages of $0.87$ to $0.89$ at $n=20{,}000$ and about $0.90$ at $n=100{,}000$; with $k_H(a)$ its coverage is between $0.925$ and $0.965$. This pattern is consistent with the variability of $\hat\Sigma_n$ with few batches, which $k_H(a)$ accounts for. Under sequential stopping the coverage of the $\chi^2$ region falls to $0.81$ and $0.85$, and the Hotelling-type quantile raises it to $0.915$ and $0.92$. The rule based on the corrected region stops about $20\%$ later, and the observed coverage of the corrected region is then $0.93$ for the Student-$t$ model and $0.925$ for the contaminated normal model, with binomial 95\% intervals of about $(0.88,0.96)$. The correction for the number of batches thus removes most of the deficit of the joint region, but under sequential stopping the observed coverage remains two to three percentage points below the nominal level. Appendix~\ref{app:covdiag} compares the variability of the chain means with the estimated Monte Carlo standard errors and examines the sensitivity of these coverages to the reference values.

\subsection{A rank-deficient design}\label{sec:crossed}

The third experiment studies the efficiency of the sampler and the Monte Carlo output analysis on a rank-deficient design covered by Theorem~\ref{thm:anydesign}. We used two crossed random factors with five levels each and four observations per cell, so that $N=100$, $q=10$ and $Z=(Z_1\ Z_2)$ has rank $r=9$, with null space spanned by $(1_5^T,-1_5^T)^T$; the intercept also lies in the column space of $Z$. The fixed effects are an intercept and a standard normal covariate, with $\beta=(1,0.5)^T$ and $\sigma_e=0.5$, and the ten random effects are Student-$t$ with $d\in\{0.5,1,2\}$ degrees of freedom, a common scale $\sigma_u=1$, and a single precision $\lambda_u$. For each $d$ we generated one data set and fitted the model with the same $d$ and the priors of Section~\ref{sec:app}. By Remark~\ref{rem:leverage}, $S_Z(c)=10^{1-2c}$, and $\min_{c}R_H(c)S_Z(c)$ equals $1.63$, $0.58$ and $0.10$ for $d=0.5$, $1$ and $2$. Theorem~\ref{thm:general} thus applies for $d=1$ and $d=2$, whereas the case $d=0.5$ is covered only by Theorem~\ref{thm:anydesign}.

We monitored $g=(\beta_1,\beta_2,\log\sigma_e,\log\sigma_u)^T$, which has posterior moments of all orders by Proposition~\ref{prop:moments}, and two bounded functions of the random effects, for which the CLT holds without moment conditions: $\arctan(u_{1}-u_{2})$, a contrast between two levels of the first factor that is identified by the data, and $\arctan(\bar u_{(1)}-\bar u_{(2)})$, the difference between the means of the two factors, which lies along $\mathcal N(Z)$ and is identified only through the prior. For each $d$, reference values came from one run of $10^6$ iterations, and we ran 200 chains of $20{,}000$ iterations after a burn-in of $1000$, started from $\beta\sim\N_2(0,4I)$, $u\sim\N_{10}(0,9I)$ and weights drawn from $H$. The split $\hat R$ statistics were computed from the first four of these chains. Table~\ref{tab:crossed} reports the results.

\begin{table}[t]
\centering\small
\caption{Crossed random factors with Student-$t$ random effects ($N=100$, $q=10$, $r=9$): Theorem~\ref{thm:general} quantity $\min_cR_H(c)S_Z(c)$; median over 200 chains of $20{,}000$ iterations of the univariate ESS per iteration and of mESS$/n$ for $g$; coverage of 95\% batch-means confidence intervals ($\lfloor n^{0.6}\rfloor$, $52$ batches) and of the joint region for $\E_\pi g$ with the $\chi^2$ and the Hotelling-type calibration. $h_1=\arctan(u_1-u_2)$ and $h_2=\arctan(\bar u_{(1)}-\bar u_{(2)})$. The simulation standard error of a coverage near $0.95$ is about $0.015$.}
\label{tab:crossed}
\begin{tabular}{lrrr}
\toprule
 & $d=0.5$ & $d=1$ & $d=2$\\
\midrule
$\min_cR_H(c)S_Z(c)$ & 1.63 & 0.58 & 0.10\\
\addlinespace
ESS$/n$: $\beta_1$, $\beta_2$ & 0.10, 0.99 & 0.33, 0.99 & 0.47, 0.98\\
\phantom{ESS$/n$:} $\log\sigma_e$, $\log\sigma_u$ & 0.86, 0.09 & 0.82, 0.14 & 0.84, 0.16\\
\phantom{ESS$/n$:} $h_1$, $h_2$ & 0.91, 0.08 & 1.03, 0.34 & 0.99, 0.60\\
mESS$/n$ for $g$ & 0.31 & 0.47 & 0.54\\
\addlinespace
Coverage: $\beta_1$, $\beta_2$ & 0.970, 0.940 & 0.905, 0.910 & 0.935, 0.950\\
\phantom{Coverage:} $\log\sigma_e$, $\log\sigma_u$ & 0.930, 0.935 & 0.960, 0.950 & 0.920, 0.950\\
\phantom{Coverage:} $h_1$, $h_2$ & 0.965, 0.935 & 0.940, 0.950 & 0.940, 0.955\\
Joint, $\chi^2$ / Hotelling & 0.920 / 0.940 & 0.875 / 0.920 & 0.920 / 0.965\\
\bottomrule
\end{tabular}
\end{table}

The chains moved freely along the null space of $Z$, with split $\hat R$ statistics at most $1.001$. Mixing is slowest for the intercept, for $\log\sigma_u$ and for $h_2$, the quantities tied to the unidentified direction, and in these three examples, each with its own data set, it is slower for smaller $d$; for $d=0.5$ the ESS per iteration of these quantities is about $0.1$, against about $0.9$ for the others. After $20{,}000$ iterations the median mESS of $g$ is then about $6200$, below $\mathrm{minESS}(4,0.05,0.05)=8431$, so the stopping rule of Section~\ref{sec:practice} would continue the run. The intervals for $d=0.5$, the case covered only by Theorem~\ref{thm:anydesign}, have coverages between $0.93$ and $0.97$, and those for $d=2$ between $0.92$ and $0.955$. For $d=1$ the intervals for $\beta_1$ and $\beta_2$ cover about $91\%$ of the time. As an exploratory follow-up to this result, we ran a further 100 chains for $d=1$: the standard deviation over chains of the estimates of $\E_\pi\beta_2$ was $1.16$ times the root mean square of their Monte Carlo standard errors, so at this run length the batch-means standard error of $\beta_2$ is somewhat too small, although its autocorrelation is negligible; for $\beta_1$ the ratio was $1.05$, and the average error of the estimates, $0.24$ Monte Carlo standard errors or about $1.8$ times the standard error of the reference value, suggests that the reference value contributes to the deficit for this component. As in the previous experiment, the Hotelling-type calibration improves the coverage of the joint region.

%% file: sec7_app.tex
\section{Application: \texorpdfstring{$\beta$}{beta}-HCG in early pregnancy}\label{sec:app}

\subsection{Data and model}
We analyse repeated measurements of the beta subunit of human chorionic gonadotropin ($\beta$-HCG) from 173 women attending a private fertility clinic in Santiago, Chile \citep{MarshallBaron2000}. Of these, 124 had a normal pregnancy and 49 an abnormal outcome. The data are sparse and irregular: the 173 women contribute $N=375$ measurements, between one and six per woman (median two), taken between 9 and 86 days. The data have been analysed with heavy-tailed and skewed error distributions \citep{DeLaCruzBranco2009}, in semiparametric classification models \citep{DeLaCruzQuintanaMuller2007}, as covariates in regression models with random effects \citep{DeLaCruz2016JMVA}, and with nonlinear mixed-effects location-scale models \citep{Alvares2025}. We fit
\begin{multline*}
\log_{10}\beta\text{-HCG}_{ij}=\beta_1+\beta_2t_{ij}+\beta_3t_{ij}^2\\+\beta_4\,\mathrm{group}_i+\beta_5\,\mathrm{group}_i\,t_{ij}+u_i+e_{ij},
\end{multline*}
where $t_{ij}$ is time in weeks, $\mathrm{group}_i$ indicates an abnormal pregnancy, $e_{ij}\iid\N(0,\lambda_e^{-1})$, and the random intercepts $u_i$ follow the scale mixtures of Table~\ref{tab:families}. Here $p=5$, $q=173$ and $Z$, the subject indicator matrix, has full column rank. We used $\beta\sim\N_5(0,1000\,I)$ and $\Ga(1,0.1)$ priors for $\lambda_e$ and $\lambda_u$. Condition \ref{C1}, $a_e>\frac12(r-N+2)=-100$ with $r=q=173$, holds, so by Theorem~\ref{thm:general} and Corollary~\ref{cor:families} the Gibbs sampler is geometrically ergodic for every model with a common scale considered below (Proposition~\ref{prop:groups} covers the models with group-specific scales of Section~\ref{sec:groupscales}), and by Proposition~\ref{prop:moments} the output analysis of Section~\ref{sec:practice} applies to $g$ in \eqref{eq:monitor}. The parameter $\beta_4$ is the difference between the groups at $t=0$, which lies before the first observation (1.3 weeks); because of the interaction, the group difference at time $t$ is $\beta_4+\beta_5t$, and we report it at four weeks as well.

\subsection{A common scale for all women}\label{sec:commonscale}
{\sloppy We fitted the normal model and 54 heavy-tailed models: Student-$t$ with $d\in\{0.5,1,1.5,2,2.5,3,\allowbreak 4,5,7,10,20\}$, slash with $\nu\in\{0.5,0.6,0.75,\allowbreak 0.9,1,1.5,2,3\}$, contaminated normal with $(\nu,\gamma)$ in $\{0.05,\allowbreak0.1,\allowbreak0.2,\allowbreak0.3\}\times\{0.05,\allowbreak0.1,\allowbreak0.2,\allowbreak0.3\}$, in $\{0.15,0.2,0.25,\allowbreak0.3,0.4\}\times\{0.01,0.02,0.03\}$ and in $\{(0.15,0.05),\allowbreak(0.25,0.05),\allowbreak(0.4,0.05)\}$, which gives 34 members, and Laplace. Each was run for $110{,}000$ iterations, the first $10{,}000$ discarded. Table~\ref{tab:waic} shows WAIC, its effective number of parameters, the log pseudo-marginal likelihood (LPML) of \citet{GeisserEddy1979}, and the posterior mean of $\beta_4$ and probability that it is positive, for a representative subset.\par}

\begin{table}[t]
\centering\small
\caption{$\beta$-HCG data, models with a common scale $\lambda_u$: WAIC, difference from the best WAIC among the 55 models fitted ($\Delta$) with its standard error computed from the subject-level contributions \citep{vehtari:gelman:gabry:2017}, effective number of parameters $p_{\mathrm{WAIC}}$, LPML, posterior mean of $\beta_4$, and $\Pr(\beta_4>0\mid y)$. Across independent runs of the same model, WAIC varied by about one unit at most; $\Delta$ is computed before rounding.}
\label{tab:waic}
\begin{tabular}{lrrrrrrr}
\toprule
Random-effects distribution & WAIC & $\Delta$ & SE & $p_{\mathrm{WAIC}}$ & LPML & $\E(\beta_4\mid y)$ & $\Pr(\beta_4>0\mid y)$\\
\midrule
Normal & 515.7 & 23.9 & 11.2 & 12.2 & -258.2 & 0.219 & 0.923 \\
\addlinespace
Student-$t$, $d=0.5$ & 544.8 & 52.9 & 3.2 & 10.3 & -272.5 & 0.467 & 0.998 \\
Student-$t$, $d=1$ & 505.8 & 14.0 & 1.3 & 10.8 & -252.9 & 0.433 & 0.997 \\
Student-$t$, $d=1.5$ & 499.8 & 8.0 & 2.3 & 10.9 & -249.8 & 0.396 & 0.993 \\
Student-$t$, $d=2$ & 499.8 & 8.0 & 3.5 & 11.3 & -250.0 & 0.363 & 0.988 \\
Student-$t$, $d=3$ & 501.8 & 10.0 & 5.2 & 11.4 & -250.9 & 0.317 & 0.977 \\
Student-$t$, $d=5$ & 505.9 & 14.1 & 7.1 & 11.7 & -253.3 & 0.269 & 0.958 \\
Student-$t$, $d=10$ & 510.1 & 18.3 & 9.1 & 11.7 & -255.0 & 0.238 & 0.939 \\
\addlinespace
Slash, $\nu=0.5$ & 508.8 & 17.0 & 2.3 & 10.5 & -254.6 & 0.423 & 0.995 \\
Slash, $\nu=0.75$ & 501.9 & 10.1 & 3.1 & 11.3 & -251.0 & 0.379 & 0.990 \\
Slash, $\nu=1$ & 502.3 & 10.5 & 4.5 & 11.5 & -251.2 & 0.337 & 0.981 \\
Slash, $\nu=2$ & 508.7 & 16.9 & 8.1 & 11.8 & -254.4 & 0.247 & 0.944 \\
\addlinespace
CN, $(\nu,\gamma)=(0.1,0.1)$ & 502.5 & 10.7 & 5.3 & 11.8 & -251.5 & 0.327 & 0.978 \\
CN, $(\nu,\gamma)=(0.1,0.05)$ & 498.2 & 6.4 & 3.9 & 11.4 & -249.1 & 0.405 & 0.993 \\
CN, $(\nu,\gamma)=(0.2,0.05)$ & 493.5 & 1.7 & 1.6 & 11.0 & -247.1 & 0.425 & 0.996 \\
CN, $(\nu,\gamma)=(0.2,0.03)$ & 491.8 & 0.0 & -- & 10.5 & -246.0 & 0.461 & 0.998 \\
CN, $(\nu,\gamma)=(0.25,0.02)$ & 492.1 & 0.3 & 1.8 & 10.3 & -246.0 & 0.474 & 0.999 \\
\addlinespace
Laplace & 501.8 & 10.0 & 6.2 & 11.2 & -251.2 & 0.298 & 0.971 \\
\bottomrule
\end{tabular}
\end{table}

\begin{figure}[t]
\centering
\includegraphics[width=\textwidth]{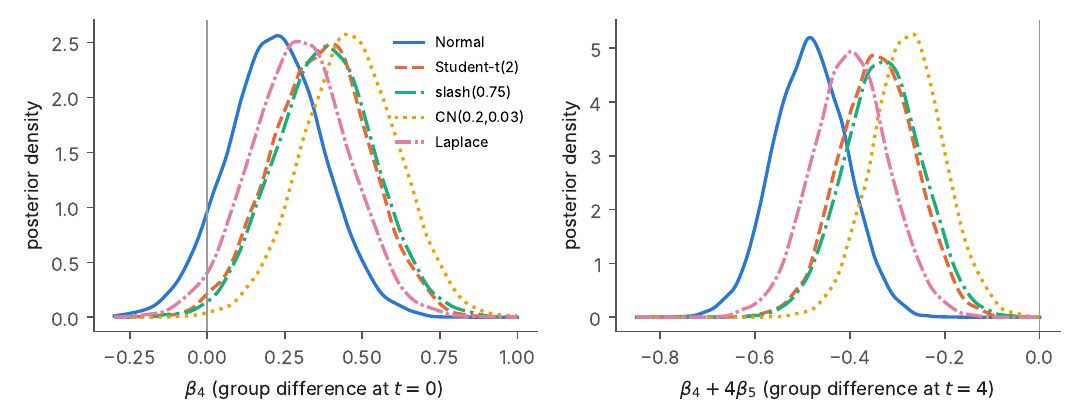}
\caption{Posterior densities of the group difference at $t=0$, $\beta_4$ (left), and at four weeks, $\beta_4+4\beta_5$ (right), under the normal model and a member close to the best of each heavy-tailed family, all with a common scale for the random effects: the members of Table~\ref{tab:post}.}
\label{fig:beta4}
\end{figure}

All heavy-tailed models in Table~\ref{tab:waic} except the Student-$t$ with $d=0.5$ improve on the normal model, by 14 to 24 WAIC units for the best member of each family. The difference between the normal model and the best model, the contaminated normal with $(\nu,\gamma)=(0.2,0.03)$, is $23.9$ with a standard error of $11.2$. Among the heavy-tailed families, the best contaminated normal improves on the best Student-$t$ ($d=1.5$) by $8.0$ units (standard error $2.3$). This comparison is somewhat optimistic for the contaminated normal, because 34 of the 54 heavy-tailed models fitted are contaminated normals on a two-dimensional grid. The improvement comes with a marked change in the estimated group effect. Under the normal model the posterior mean of $\beta_4$ is $0.22$ and the 95\% credible interval, $(-0.08,0.52)$, contains zero. Under the best contaminated normal model the posterior mean doubles to $0.46$, with interval $(0.16,0.76)$. Table~\ref{tab:post} gives posterior summaries for the best model in each family, and Figure~\ref{fig:beta4} the posterior densities of $\beta_4$. The other coefficients are stable across models; the scale $\sigma_u$ is not comparable across families, because it plays a different role in each. The change in $\beta_4$ concerns an extrapolation to $t=0$. At four weeks the posterior mean of the group difference $\beta_4+4\beta_5$ is $-0.48$, with 95\% interval $(-0.63,-0.33)$, under the normal model and $-0.28$, with interval $(-0.43,-0.13)$, under the best contaminated normal model, so both models find lower $\beta$-HCG levels in abnormal pregnancies from about four weeks on. Across the members of Table~\ref{tab:post} its posterior mean ranges from $-0.48$ to $-0.28$, a spread comparable to that of $\beta_4$, but every 95\% interval excludes zero, so the qualitative conclusion at an observed time does not depend on the random-effects distribution. Figure~\ref{fig:beta4} shows both contrasts.

\begin{table}[t]
\centering\small
\caption{$\beta$-HCG data, models with a common scale: posterior means (standard deviations) for a model close to the best in each family (for the Student-$t$, $d=2$, within $0.1$ WAIC units of the best, $d=1.5$), based on four chains of $55{,}000$ iterations after burn-in, together with output-analysis summaries. mESS$/n$ and the number of iterations needed to reach $\mathrm{minESS}=8769$, estimated as $n\,\mathrm{minESS}/\mathrm{mESS}$ with $n=55{,}000$ and batch size $\lfloor n^{0.6}\rfloor$, refer to a single chain and to $g$ in \eqref{eq:monitor}, and so does WAIC, which differs from the value in Table~\ref{tab:waic}, obtained from an independent run, by at most one unit; $\hat R$ is the split potential scale reduction factor over the four chains.}
\label{tab:post}
\begin{tabular}{lrrrrr}
\toprule
 & Normal & $t_2$ & Slash(0.75) & CN(0.2,0.03) & Laplace\\
\midrule
$\beta_1$ & 0.467 (0.149) & 0.427 (0.137) & 0.423 (0.137) & 0.419 (0.133) & 0.437 (0.140) \\
$\beta_2$ & 1.118 (0.052) & 1.135 (0.048) & 1.136 (0.048) & 1.138 (0.047) & 1.131 (0.050) \\
$\beta_3$ & -0.069 (0.004) & -0.071 (0.004) & -0.071 (0.004) & -0.071 (0.004) & -0.070 (0.004) \\
$\beta_4$ & 0.218 (0.153) & 0.364 (0.159) & 0.380 (0.161) & 0.463 (0.155) & 0.298 (0.156) \\
$\beta_5$ & -0.175 (0.028) & -0.176 (0.027) & -0.177 (0.028) & -0.185 (0.027) & -0.174 (0.028) \\
$\beta_4+4\beta_5$ & -0.484 (0.076) & -0.341 (0.081) & -0.330 (0.082) & -0.278 (0.076) & -0.399 (0.080) \\
$\sigma_e$ & 0.381 (0.019) & 0.365 (0.017) & 0.363 (0.017) & 0.362 (0.017) & 0.366 (0.018) \\
$\sigma_u$ & 0.333 (0.033) & 0.187 (0.023) & 0.140 (0.017) & 0.157 (0.019) & 0.250 (0.027) \\
\midrule
WAIC & 515.7 & 500.1 & 501.6 & 492.1 & 502.8 \\
$\Pr(\beta_4>0\mid y)$ & 0.922 & 0.988 & 0.990 & 0.998 & 0.971 \\
mESS / $n$ & 0.87 & 0.51 & 0.51 & 0.52 & 0.56 \\
Iterations for minESS & 10\,043 & 17\,143 & 17\,201 & 16\,890 & 15\,574 \\
$\max\hat R$ & 1.0001 & 1.0001 & 1.0001 & 1.0001 & 1.0002 \\
\bottomrule
\end{tabular}
\end{table}

\subsection{What the latent weights indicate}
\begin{figure}[t]
\centering
\includegraphics[width=\textwidth]{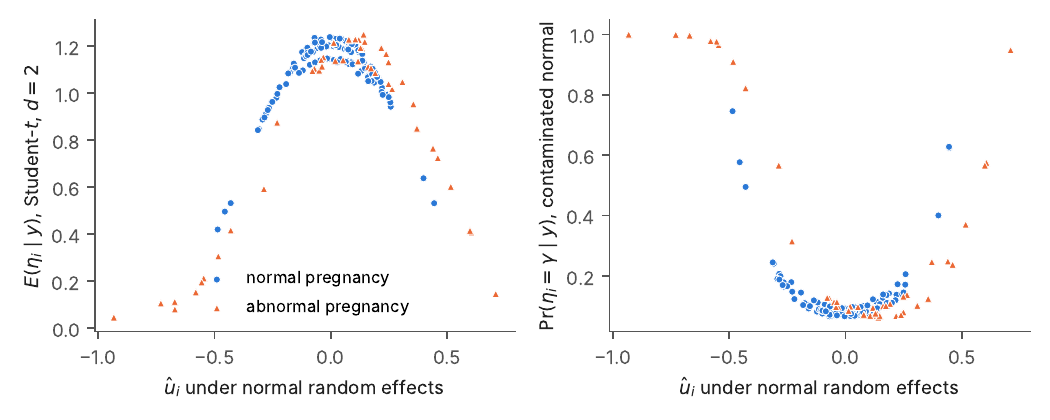}
\caption{Left: posterior mean of the latent weight $\eta_i$ under the Student-$t_2$ model. Right: posterior probability that woman $i$ belongs to the contaminated component under the CN(0.2,0.03) model. Both are plotted against the posterior mean of $u_i$ under the normal model; circles are normal and triangles abnormal pregnancies.}
\label{fig:weights}
\end{figure}

\begin{figure}[t]
\centering
\includegraphics[width=\textwidth]{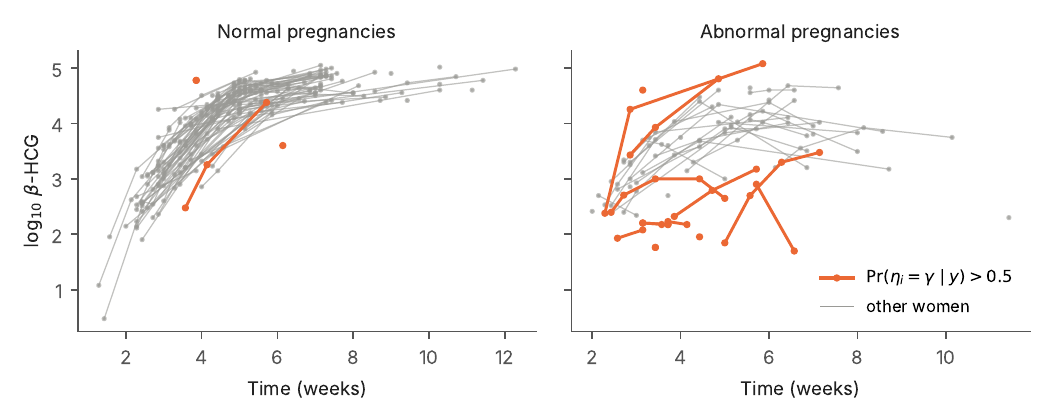}
\caption{Observed $\log_{10}\beta$-HCG profiles by group. Highlighted profiles correspond to the 19 women with posterior probability above $0.5$ of belonging to the contaminated component under the CN(0.2,0.03) model.}
\label{fig:profiles}
\end{figure}

The latent weights make it possible to see which women drive the improvement. Figure~\ref{fig:weights} plots, for each woman, the posterior mean weight $\E(\eta_i\mid y)$ under the $t_2$ model and the posterior probability of belonging to the contaminated component, $\Pr(\eta_i=\gamma\mid y)$, under the CN(0.2,0.03) model, against her predicted random effect under the normal model. Women with extreme random effects are downweighted, as intended. The posterior probabilities $\Pr(\eta_i=\gamma\mid y)$ are means of bounded functions and are covered by the CLT; the posterior means of the weights are reported descriptively. More informative is who these women are: of the 19 women with $\Pr(\eta_i=\gamma\mid y)>0.5$, 15 had an abnormal pregnancy. That is 31\% of the abnormal group, against 3\% of the normal group. Fourteen of these 19 women have a negative predicted random effect under the normal model, that is, levels below those of the population curve. Figure~\ref{fig:profiles} shows their profiles. The same pattern is visible in the predicted random effects of the normal model, whose standard deviation is $0.18$ in the normal and $0.39$ in the abnormal group; the weights summarize it in a form that is directly tied to the model.

\subsection{Group-specific scales}\label{sec:groupscales}
The pattern of the latent weights suggests that the departure from the normal model is not so much a heavy tail common to all women as a larger between-subject variability among abnormal pregnancies, which is consistent with the location-scale analysis of \citet{Alvares2025}; see also \citet{hede:merm:demi:2008} for mixed-effects location-scale models. We therefore refitted a selection of the models allowing a separate scale for each group, as in \eqref{eq:groupscale}. Proposition~\ref{prop:groups} guarantees that these chains are geometrically ergodic, and Proposition~\ref{prop:moments} applies to them as well. Table~\ref{tab:hetero} shows the results. For these fits we used a smaller set of members: Student-$t$ with $d\in\{2,4,8,16\}$, slash with $\nu\in\{1,4\}$, the contaminated normal with $(\nu,\gamma)\in\{(0.05,0.25),(0.1,0.1)\}$, and the Laplace law. Members lighter than those of the common-scale grid, $d=8$, $d=16$, $\nu=4$ and $(\nu,\gamma)=(0.05,0.25)$, were included because with group-specific scales the fit improves as the tails become lighter. The model with group-specific scales and normal random effects has a WAIC of $464.9$, an improvement of $27$ units (standard error $13$) over the best common-scale model, and the estimated standard deviation of the random intercepts is $0.18$ for normal and $0.62$ for abnormal pregnancies. Once the scale is allowed to differ between groups, heavy tails no longer help. The lighter-tailed members of each family are within $1.6$ WAIC units of the normal model (less than two standard errors), and the heavier ones are worse, for example by $10.6$ units (standard error $4.9$) for the Student-$t$ with $d=2$. The posterior of $\beta_4$ returns to a mean of $0.20$, with interval $(-0.14,0.53)$ and $\Pr(\beta_4>0\mid y)=0.87$. Among models with similar WAIC, the extrapolated difference at $t=0$ still depends on the mixing distribution; for the Laplace model ($\Delta=2.7$, standard error $4.0$), $\Pr(\beta_4>0\mid y)=0.97$. The group difference at four weeks, which refers to an observed time, is stable: its posterior mean ranges from $-0.49$ to $-0.38$ across the models of Table~\ref{tab:hetero}, with 95\% intervals that exclude zero in every case, for example $(-0.69,-0.29)$ under the normal and $(-0.58,-0.20)$ under the Laplace model. Adding a group-by-$t^2$ term to the normal model with group-specific scales gives a coefficient with posterior mean $0.00$ and interval $(-0.02,0.02)$, and increases WAIC by $4.5$ (standard error $2.2$), so this extension of the mean structure receives no support in the comparison.

\begin{table}[t]
\centering\footnotesize\setlength{\tabcolsep}{3.5pt}
\caption{$\beta$-HCG data, models with group-specific scales \eqref{eq:groupscale}: WAIC, difference from the best ($\Delta$) with its standard error, posterior means of $\sigma_{u,k}=\lambda_{u,k}^{-1/2}$ for normal ($k=1$) and abnormal ($k=2$) pregnancies, posterior mean and probability of positivity of $\beta_4$, and posterior mean and 95\% credible interval of the group difference at four weeks, $\beta_4+4\beta_5$.}
\label{tab:hetero}
\begin{tabular}{lrrrrrrrrr}
\toprule
Random effects & WAIC & $\Delta$ & SE & $\sigma_{u,1}$ & $\sigma_{u,2}$ & $\E(\beta_4\mid y)$ & $\Pr(\beta_4>0\mid y)$ & $\E(\beta_4+4\beta_5\mid y)$ & 95\% interval\\
\midrule
Normal & 464.9 & 0.0 & -- & 0.178 & 0.621 & 0.196 & 0.873 & -0.492 & (-0.695, -0.289) \\
Student-$t$, $d=16$ & 465.7 & 0.8 & 0.6 & 0.171 & 0.586 & 0.210 & 0.889 & -0.477 & (-0.681, -0.272) \\
Student-$t$, $d=8$ & 466.4 & 1.6 & 1.3 & 0.164 & 0.555 & 0.224 & 0.903 & -0.461 & (-0.667, -0.258) \\
Student-$t$, $d=4$ & 468.5 & 3.6 & 2.7 & 0.152 & 0.497 & 0.256 & 0.930 & -0.431 & (-0.636, -0.231) \\
Student-$t$, $d=2$ & 475.5 & 10.6 & 4.9 & 0.132 & 0.401 & 0.313 & 0.966 & -0.382 & (-0.581, -0.190) \\
Slash, $\nu=4$ & 465.6 & 0.7 & 0.6 & 0.162 & 0.544 & 0.201 & 0.878 & -0.486 & (-0.691, -0.281) \\
Slash, $\nu=1$ & 475.8 & 10.9 & 3.6 & 0.123 & 0.359 & 0.264 & 0.935 & -0.426 & (-0.635, -0.223) \\
CN$(0.05,0.25)$ & 466.0 & 1.1 & 0.6 & 0.171 & 0.595 & 0.200 & 0.877 & -0.485 & (-0.690, -0.281) \\
CN$(0.1,0.1)$ & 470.3 & 5.5 & 1.9 & 0.156 & 0.541 & 0.218 & 0.898 & -0.467 & (-0.673, -0.261) \\
Laplace & 467.6 & 2.7 & 4.0 & 0.148 & 0.471 & 0.311 & 0.965 & -0.384 & (-0.575, -0.196) \\
\bottomrule
\end{tabular}
\end{table}

The analysis therefore leads to a different conclusion from the one suggested by the common-scale models. Heavy-tailed random effects fit better than normal ones when a single scale is imposed, and with that restriction they change the inference about the group difference. Their latent weights, however, point to group heteroscedasticity as the underlying feature, and a normal model with group-specific variances fits better than any of the common-scale models. Under it, the evidence for a difference between groups at time zero is weak, whereas all models agree that abnormal pregnancies have lower levels from about four weeks on. The downweighted women have mostly low levels, which suggests some asymmetry in the random-effects distribution of the abnormal group; skewed random effects \citep{Lachos2010,Schumacher2021} would be a natural next step. In this example the scale-mixture models are thus most useful as a diagnostic.

\subsection{Sensitivity to the prior on the scale}\label{sec:priorsens}
This subsection concerns the models with a common scale of Section~\ref{sec:commonscale}. In those models, the same $\Ga(1,0.1)$ prior on $\lambda_u$ does not induce the same prior on a comparable measure of dispersion in every family, because $\sigma_u$ plays a different role in each, and for the Student-$t$ with $d\le2$ and the slash with $\nu\le1$ the random effects have no variance. As a common scale we use the median absolute deviation of the random-effects distribution, $\tau=m_H\sigma_u$, where $m_H$ is the median of $|\xi|\eta^{-1/2}$ with $\xi\sim\N(0,1)$ independent of $\eta\sim H$. Thus $m_H=0.674$ for the normal law, $0.873$ and $0.817$ for the Student-$t$ with $d=1.5$ and $d=2$, $1.176$ for the slash law with $\nu=0.75$, $0.835$ for the contaminated normal with $(\nu,\gamma)=(0.2,0.03)$ and $0.693$ for the Laplace law. A $\Ga(1,b)$ prior on $\tau^{-2}$ is a $\Ga(1,b/m_H^2)$ prior on $\lambda_u$, so the sampler and the theory are unchanged. We refitted the best member of each family, and the Student-$t$ with $d=2$, with $b=b_0m_N^2$, $m_N=0.674$ and $b_0\in\{0.01,0.1,1\}$; for the normal model these are the $\Ga(1,b_0)$ priors on $\lambda_u$, and $b_0=0.1$ gives the prior used above. Each fit used one chain of $45{,}000$ iterations, the first $5000$ discarded. Table~\ref{tab:prior} shows the results. With $b_0=0.01$ and $b_0=0.1$, WAIC changes by less than one unit and the posterior means of $\beta_4$ and $\beta_4+4\beta_5$ by at most $0.017$. The prior with $b_0=1$ has a prior median of $\tau$ of about $0.81$, well above the posterior values, which lie between $0.11$ and $0.25$. It increases WAIC by $1.0$ for the normal model and by $3.6$ to $11.4$ units for the heavy-tailed ones, so the improvement over the normal model under these priors shrinks from between $13$ and $24$ units to between $7.5$ and $13.7$ units; the contaminated normal remains the best model. Across the three priors, the posterior means of $\beta_4$ and $\beta_4+4\beta_5$ vary by at most $0.034$ and $0.030$, respectively. The comparison of families is therefore somewhat sensitive to an informative prior on the common scale, whereas the posterior means of the group contrasts, and the probabilities $\Pr(\beta_4>0\mid y)$, which change by at most $0.005$, are relatively stable.

\begin{table}[t]
\centering\footnotesize\setlength{\tabcolsep}{3pt}
\caption{$\beta$-HCG data, models with a common scale: sensitivity to a $\Ga(1,b_0m_N^2)$ prior on $\tau^{-2}$, where $\tau$ is the median absolute deviation of the random-effects distribution and $m_N=0.674$; $b_0=0.1$ corresponds to the prior of Tables~\ref{tab:waic} and~\ref{tab:post} for the normal model. WAIC and posterior means of $\tau$, $\beta_4$ and $\beta_4+4\beta_5$, from one chain of $40{,}000$ iterations after burn-in.}
\label{tab:prior}
\begin{tabular}{lrrrrrrrrrrrr}
\toprule
 & \multicolumn{3}{c}{WAIC} & \multicolumn{3}{c}{$\E(\tau\mid y)$} & \multicolumn{3}{c}{$\E(\beta_4\mid y)$} & \multicolumn{3}{c}{$\E(\beta_4+4\beta_5\mid y)$}\\
\cmidrule(lr){2-4}\cmidrule(lr){5-7}\cmidrule(lr){8-10}\cmidrule(lr){11-13}
$b_0$ & 0.01 & 0.1 & 1 & 0.01 & 0.1 & 1 & 0.01 & 0.1 & 1 & 0.01 & 0.1 & 1\\
\midrule
Normal & 515.8 & 515.6 & 516.6 & 0.221 & 0.225 & 0.253 & 0.219 & 0.219 & 0.221 & -0.483 & -0.483 & -0.486 \\
Student-$t$, $d=1.5$ & 499.8 & 499.5 & 509.1 & 0.114 & 0.131 & 0.195 & 0.399 & 0.398 & 0.387 & -0.316 & -0.318 & -0.331 \\
Student-$t$, $d=2$ & 500.0 & 499.6 & 506.2 & 0.136 & 0.148 & 0.202 & 0.363 & 0.363 & 0.357 & -0.339 & -0.341 & -0.351 \\
Slash, $\nu=0.75$ & 501.3 & 500.6 & 507.9 & 0.128 & 0.144 & 0.203 & 0.388 & 0.386 & 0.367 & -0.320 & -0.324 & -0.342 \\
CN, $(0.2,0.03)$ & 491.6 & 491.5 & 502.9 & 0.112 & 0.125 & 0.185 & 0.437 & 0.454 & 0.471 & -0.296 & -0.284 & -0.266 \\
Laplace & 502.8 & 502.6 & 506.2 & 0.167 & 0.173 & 0.211 & 0.293 & 0.296 & 0.314 & -0.401 & -0.400 & -0.392 \\
\bottomrule
\end{tabular}
\end{table}

\subsection{Monte Carlo accuracy}\label{sec:appoutput}
\begin{figure}[t]
\centering
\includegraphics[width=0.62\textwidth]{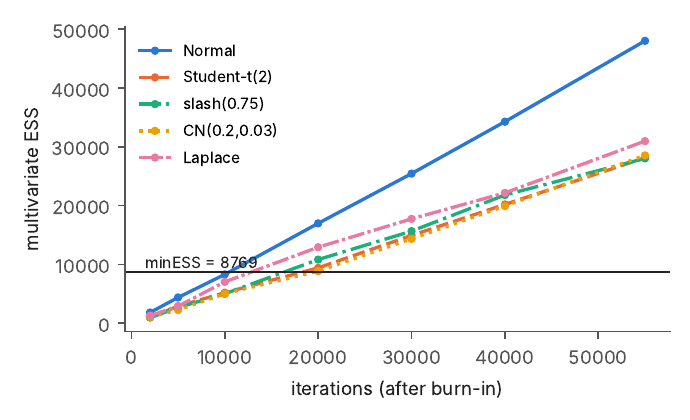}
\caption{Multivariate effective sample size of $g=(\beta^T,\log\sigma_e,\log\sigma_u)^T$ as a function of the number of iterations, for one chain of each model in Table~\ref{tab:post}. The horizontal line is $\mathrm{minESS}(7,0.05,0.05)$.}
\label{fig:mess}
\end{figure}

For each model in Table~\ref{tab:post} we ran four chains from dispersed starting values. The split $\hat R$ statistics are at most $1.0002$. Because the chains are geometrically ergodic and $g$ has posterior moments of all orders, the batch-means standard errors with batch size $\lfloor n^{0.6}\rfloor$ and the multivariate effective sample size are strongly consistent (Section~\ref{sec:practice}). Figure~\ref{fig:mess} shows how mESS grows with the run length. The heavy-tailed models are less efficient than the normal one, with mESS per iteration of about $0.51$ to $0.56$ against $0.87$. Heuristically, small values of $\eta_i$ weaken the link between $u_i$ and $\lambda_u$, and the latent weights add autocorrelation; geometric ergodicity does not quantify this effect. Even so, extrapolating from runs of $55{,}000$ iterations, all models reach $\mathrm{minESS}=8769$ in fewer than $18{,}000$ iterations. Figure~\ref{fig:mess} confirms this at $20{,}000$ iterations. On a laptop with a $2$ GHz processor, $20{,}000$ iterations take about $1.5$ to $5$ seconds. The Monte Carlo standard errors of the posterior means of $\beta$, $\log\sigma_e$ and $\log\sigma_u$ are at most $0.0017$ after $55{,}000$ iterations, so the posterior means of $\beta$ in Table~\ref{tab:post} are accurate to about $\pm0.003$. For the scales on their natural scale, Section~\ref{sec:practice} gives posterior moments of $\sigma_e$ of every order below $2a_e+N=377$, so the CLT also applies to $\sigma_e$, whereas the bound for $\sigma_u$, of order below $2a_u=2$, is not sufficient for a CLT; this is why $g$ contains $\log\sigma_u$. Since $\sigma_e$ has posterior moments of order larger than four, the CLT also applies to $(\sigma_e,\sigma_e^2)$, and the delta method gives Monte Carlo standard errors for the posterior standard deviation of $\sigma_e$. For $\sigma_u$ the available bounds do not cover its posterior standard deviation, and WAIC is not covered by this theory either; their stability was checked across the four chains.

For the models with group-specific scales of Table~\ref{tab:hetero}, single chains of $100{,}000$ iterations gave mESS per iteration between $0.49$ and $0.69$, that is, an mESS between $49{,}000$ and $69{,}000$, well above $\mathrm{minESS}(8,0.05,0.05)=8804$, and Monte Carlo standard errors at most $0.0018$ for the components of $g$. These are the posterior summaries on which the comparisons of the group effect rest; for WAIC and LPML, which are not covered by Proposition~\ref{prop:moments}, we relied on the agreement between independent runs.

%% file: sec8_disc.tex
\section{Discussion}\label{sec:disc}

The three-block Gibbs sampler for Bayesian linear mixed models with scale mixtures of normal random effects is geometrically ergodic under mild conditions on the mixing distribution. The conditions hold for all the families commonly used for random effects and for every random-effects design matrix, full rank or not; for these families, the rank of the design enters only the conditions on the shapes of the gamma priors. Two devices make this possible: the coordinatewise bound of Lemma~\ref{lem:trace}(a), which separates the row space of the design from its null space and leads to the condition $R_H(c)S_Z(c)<1$ of Theorem~\ref{thm:general}, and the equivariance of the sampler under rescaling of the columns of the design, which allows the null space to be concentrated on a few coordinates (Theorem~\ref{thm:anydesign}).

Combined with posterior moment bounds that hold for every mixing distribution, geometric ergodicity justifies Monte Carlo standard errors and fixed-volume stopping rules for the regression coefficients and the log-variance components. The justification is asymptotic. The drift function and the minorization condition that holds on its sublevel sets (Lemma~\ref{lem:topology} and \citealp[Theorem~5.5.7]{meyn:twee:1993}) could in principle be combined into an explicit bound on the total variation distance \citep{rose:1995}, but such bounds are very conservative in high dimensions \citep{qin:hobe:2021}, and here the marginal chain has dimension $2q+p$, which is $351$ in the application. In the simulations, componentwise intervals were close to nominal at the run lengths at which the stopping rule terminates; joint regions required a correction for the number of batches, and under sequential stopping their observed coverage remained two to three percentage points below the nominal level, a difference of the order of the simulation error. A finite-sample calibration of fixed-volume stopping rules would complement the asymptotic theory.

In the application, the heavy-tailed models were most useful as a diagnostic: with a common scale they improved the fit and changed the inference about the group difference at baseline, but their latent weights pointed to group heteroscedasticity, and a normal model with group-specific scales fitted better than any common-scale model.

Several questions remain open. Our conditions are sufficient but not necessary, and a sharper characterization of geometric ergodicity would settle whether a mixing distribution with $m_{1/2}<\infty$ but $R_H(c)\ge1$ for every $c\in(0,1]$ can produce a chain that is not geometrically ergodic when $Z$ is rank deficient, whether \ref{C2} can be weakened, and what happens when \ref{H1} fails, either because $m_{c+1/2}=\infty$ for every $c>0$ or because $m_{1/2}=\infty$, as for variance-gamma laws with shape at most $\frac12$ and horseshoe-type mixing distributions (Remark~\ref{rem:boundary}). For regression models with heavy-tailed errors and proper priors, \citet{hamu:2024} removed the conditions on the mixing density near the origin, and a similar simplification may be possible here. The model also has one latent weight per random effect and a single scale, or scales indexed by a known grouping. In robust random-coefficient models the coordinates of a subject's random-effect vector usually share one weight and have an unstructured covariance matrix with a Wishart prior \citep{Pinheiro2001,Lachos2010}; the tilted family then becomes $\eta^{k/2}e^{-s\eta}H(d\eta)$, with $k$ the dimension of the random-effect vector, and the scale no longer cancels in the drift as it does here. Several variance components with a rank-deficient design, tail parameters with a prior on a finite grid, and heavy-tailed errors, which add a second set of weights to the conditional variance of the regression parameters, require block versions of the drift argument. A continuous prior on the tail parameter breaks the conditional conjugacy and may require an additional nonconjugate update, whose analysis calls for different tools \citep{qin:jone:2022}. Finally, quantitative bounds that remain informative as the number of subjects grows, in the spirit of \citet{jin:hobe:2022a} and \citet{asco:zane:2024}, would clarify how heavy tails affect the efficiency of the sampler.

%% file: appendix.tex
\section{An auxiliary result}\label{app:KH}

The proof of Lemma~\ref{lem:means} relies on the following uniform bound for weighted least squares projections, due to \citet{Stewart1989} and \citet{OLeary1990}. It plays the role that Lemma~3 in Appendix~C of \citet{khar:hobe:2011} plays in earlier analyses of Gibbs samplers for linear mixed models \citep{roma:hobe:2012,roma:hobe:2014}, and has the advantage of applying directly to the conditional mean of $\theta$.

\begin{lemma}\label{lem:stewart}
Let $A$ be a real $n\times m$ matrix of full column rank. Then
\[
\chi_A=\sup_{D}\norm{A(A^TDA)^{-1}A^TD}_2<\infty,
\]
where the supremum is over all $n\times n$ diagonal matrices $D$ with positive diagonal entries.
\end{lemma}

The matrix $A(A^TDA)^{-1}A^TD$ is the oblique projection that maps $b$ to the fitted values $A\hat\theta_D$ of the weighted least squares problem $\min_\theta\norm{D^{1/2}(b-A\theta)}^2$. Lemma~\ref{lem:stewart} states that these fitted values are bounded by a multiple of $\norm{b}$, uniformly in the weights. \citet{Stewart1989} proves the bound and \citet{OLeary1990} identifies $\chi_A$ in terms of the principal angles between the range of $A$ and the coordinate subspaces.

\section{Verification of the implementation}\label{app:code}

The implementation was checked in three ways. First, for each family in Table~\ref{tab:families} and several values of $s$, sample moments of $\eta$ and $\log\eta$ from $2\times10^5$ draws of $H_s$ agreed with the corresponding integrals computed by adaptive quadrature within Monte Carlo error.

Second, we applied the joint distribution test of \citet{Geweke2004} to a small model with $N=8$, $p=2$, $q=4$, $\beta\sim\N_2(0,I)$ and $\lambda_e,\lambda_u\sim\Ga(3,2)$, for the Student-$t$ ($d=3$), slash ($\nu=1.2$), contaminated normal ($(\nu,\gamma)=(0.2,0.1)$), normal inverse Gaussian ($\eta\sim\GIG(\frac12,1,1)$), variance-gamma ($\eta\sim\mathrm{IG}(2,1)$) and Laplace families, each with a design for which $Z^TZ$ is diagonal and with a non-diagonal design. With these priors $\E[\lambda_u^{-1}]=1$ and $\E[\eta_i^{-1}]<\infty$ for all six mixing distributions (for the slash law, $\E[\eta^{-1}]=\nu/(\nu-1)=6$), so all the monitored functions have finite prior second moments. In all twelve configurations the successive-conditional simulator reproduced the prior moments of $\beta$, $u$, $\lambda_e$, $\lambda_u$ and $\log\eta_1$ within Monte Carlo error. For six of them (contaminated normal and Laplace with the non-diagonal design, normal inverse Gaussian and variance-gamma with both designs) we used runs of $3\times10^5$ iterations and fourteen moments: the means and second moments of $\beta_1$, $\beta_2$, $\lambda_e$, $\lambda_u$ and $\log\eta_1$, and the means and moments of order $\frac12$ of $|u_1|$ and $|u_4|$. Standardized by batch-means standard errors, which serve here only as a practical scale for an empirical check, the 84 differences had a largest absolute value of $2.35$, and two of them exceeded $2$, against about four expected by chance.

Third, for Student-$t$ random effects with $d=1$, for which the successive-conditional simulator mixes too slowly to be informative, the Gibbs sampler agreed with a long random-walk Metropolis run targeting the posterior with the weights integrated out, for all parameters within twice the Monte Carlo standard error.

The subject-level marginal likelihoods used for WAIC and LPML in Sections~\ref{sec:sim} and~\ref{sec:app} integrate out $u_i$ and $\eta_i$. Given $(\beta,\lambda_e,\lambda_u)$ and $\eta_i$, the $n_i$ observations of subject $i$ are normal with covariance $\lambda_e^{-1}I+(\lambda_u\eta_i)^{-1}J$, where $J$ is the matrix of ones. Its determinant and quadratic form have closed forms, and the integral over $\eta_i$ is computed exactly for the contaminated normal and by numerical quadrature otherwise: a Gauss--Jacobi rule with weight $\eta^{\nu-1/2}$ on $(0,1)$ for the slash, which absorbs the behaviour of the subject likelihood near $\eta=0$, and a trapezoidal rule in $\log\eta$ for the other families. Compared with adaptive quadrature at several parameter values, the relative error of the subject likelihoods was below $10^{-10}$.

\section{Diagnostics for the coverage study}\label{app:covdiag}

For the chains of length $20{,}000$ of Section~\ref{sec:coverage} we compared the covariance matrix of the 200 chain means with the average of the estimates $\hat\Sigma_n/n$. For both models the standard deviations of the chain means are within $8\%$ of the root mean square of the Monte Carlo standard errors. The eigenvalues of the first matrix relative to the second lie between $0.74$ and $1.27$ for the Student-$t$ model and between $0.81$ and $1.25$ for the contaminated normal model, with averages $0.99$ and $1.00$; part of their spread may be due to the estimation error of a covariance matrix from 200 chains. The two matrices thus agree overall, although the comparison cannot exclude moderate discrepancies in particular directions.

For the rule based on the corrected region, with stopping times $T_r$, we compared the covariance matrix of the stopped means $\bar g_{T_r}$ with $R^{-1}\sum_{r=1}^R\hat\Sigma_{T_r}/T_r$, $R=200$. The standard deviations of the stopped means exceed the root mean square of the Monte Carlo standard errors by up to $13\%$, and the relative eigenvalues lie between $0.79$ and $1.34$ for the Student-$t$ model and between $0.68$ and $1.50$ for the contaminated normal model, with averages $1.06$. At the stopping time, the estimated covariance of the mean therefore tends to understate the observed variability of the stopped means. This is compatible with a selection effect, since a rule that stops when the estimated mESS first crosses a threshold favours runs in which $\hat\Sigma_n$ is small, although the experiment does not isolate that mechanism.

The averages of the chain means of length $20{,}000$ differ from the reference values by up to $0.33$ Monte Carlo standard errors for $\beta_1$, $\beta_2$ and $\beta_3$, that is, by $1.5$ to $3$ times the standard error of the reference values. This difference may come from the reference values, from the initial transient, or from both. As a descriptive sensitivity check, we kept the intervals unchanged and recomputed their coverage with the average of all the runs in place of the independent reference value; the coverages change by at most $0.035$. Since each run then contributes to the value against which its interval is evaluated, and a transient bias shared by all runs would not be detected, this check is not an independent assessment of coverage, and the coverages of Table~\ref{tab:coverage} use the independent reference run.